\documentclass[floats, aps, showpacs, twocolumn, superscriptaddress, pre,
    reprint, floatfix]{revtex4-1}

\usepackage{graphicx}
\graphicspath{{figures/}}
\usepackage{amsmath, amssymb, amstext, amsthm, amsfonts}
\usepackage{bbm}
\usepackage[multidot]{grffile}
\usepackage{color}

\definecolor{darkblue}{rgb}{0,0,.55}
\definecolor{redcolor}{rgb}{0.7,0.,0.}

\usepackage[colorlinks,bookmarks=false,
            citecolor=darkblue,
            linkcolor=darkblue,
            urlcolor=darkblue]{hyperref}
\usepackage[all]{hypcap}
\usepackage{dsfont}

\newcommand{\ue}{\ensuremath{\text{e}}}
\newcommand{\ui}{\ensuremath{\text{i}}}
\newcommand{\uc}{\ensuremath{\text{c}}}
\newcommand{\ud}{\ensuremath{\text{d}}}

\newcommand{\muinv}{\ensuremath{\mu_{\text{inv}}}}
\newcommand{\ginv}{\ensuremath{\gamma_{\text{inv}}}}
\newcommand{\muxi}{\ensuremath{\mu_\xi}}
\newcommand{\tilmuxi}{\ensuremath{\tilde{\mu}_\xi}}
\newcommand{\gxi}{\ensuremath{\gamma_{\xi}}}

\newcommand{\munat}{\ensuremath{\mu_{\text{nat}}}}

\newcommand{\gnat}{\ensuremath{\gamma_{\text{nat}}}}
\newcommand{\gmax}{\ensuremath{\gamma_{\text{max}}}}
\newcommand{\gmin}{\ensuremath{\gamma_{\text{min}}}}
\newcommand{\muL}{\ensuremath{\mu_{\text{L}}}}
\newcommand{\Map}{\ensuremath{M}}
\newcommand{\Mapcls}{\ensuremath{{M}}}
\newcommand{\QMap}{\ensuremath{\mathcal{U}}}
\newcommand{\QMapcls}{\ensuremath{\tilde{\mathcal{U}}}}

\newcommand{\MMap}{\ensuremath{\mathcal{M}}}
\newcommand{\refl}{R}
\newcommand{\Qrefl}{\mathcal{R}}
\newcommand{\Qtrans}{\mathcal{T}}
\newcommand{\proj}{\mathcal{P}}
\newcommand{\PS}{\ensuremath{\Gamma}}

\newcommand{\alphaOmega}{\ensuremath{R_\Omega}}

\newcommand{\hus}{\mathcal{H}}
\newcommand{\husavg}{\langle\mathcal{H}\rangle_\gamma}
\newcommand{\Dim}[1][1]{\ensuremath{D_{#1}}}

\newcommand{\jsd}{\ensuremath{d_{\text{JS}}}}

\begin{document}

\title{Structure of resonance eigenfunctions for chaotic
     systems with partial escape}

\newcommand{\affilTUD}{
    Technische Universit\"at Dresden,
    Institut f\"ur Theoretische Physik and Center for Dynamics,
    01062 Dresden, Germany}
\newcommand{\affilMPI}{
    Max-Planck-Institut f\"ur Physik komplexer Systeme,
    N\"othnitzer Stra\ss{}e 38, 01187 Dresden, Germany}
\newcommand{\affilSYD}{
    School of Mathematics and Statistics,
    University of Sydney, 2006 NSW, Australia}

\author{Konstantin Clau\ss}
\affiliation{\affilTUD}

\author{Eduardo G. Altmann}
\affiliation{\affilSYD}

\author{Arnd B\"acker}
\affiliation{\affilTUD}
\affiliation{\affilMPI}

\author{Roland Ketzmerick}
\affiliation{\affilTUD}
\affiliation{\affilMPI}

\date{\today}

\begin{abstract}
Physical systems are often neither completely closed nor completely open, but 
instead they are best described by dynamical systems with partial escape or 
absorption. In this paper we introduce classical measures that explain the main 
properties of resonance eigenfunctions of chaotic quantum systems with partial 
escape. We construct a family of conditionally-invariant measures with varying 
decay rates by interpolating between the natural measures of the forward and 
backward dynamics. Numerical simulations in a representative system show that 
our classical measures correctly describe the main features of the quantum 
eigenfunctions: their multi-fractal phase space distribution, their product 
structure along stable/unstable directions, and their dependence on the decay 
rate. The (Jensen-Shannon) distance between classical and quantum measures 
goes to zero in the semiclassical limit for long- and short-lived 
eigenfunctions, while it remains finite for intermediate cases.
\end{abstract}

\maketitle

\noindent

\section{Introduction.}%

Quantum chaos aims to relate observations in quantum systems
to properties of the classical dynamics.
Particularly interesting are the implications of the classically chaotic 
dynamics 
on the statistics of eigenvalues 
\cite{BohGiaSch1984,Ber1985,SieRic2001,MueHeuBraHaaAlt2004} and the
distribution of eigenfunctions \cite{Vor1979,Ber1977b,Ber1983}.
In closed chaotic systems typical trajectories explore the
phase space uniformly with respect to the Liouville measure.
The quantum ergodicity theorem states that
semiclassically almost all chaotic eigenfunctions
converge to this measure 
\cite{Shn1974,CdV1985,Zel1987,ZelZwo1996,NonVor1998,BaeSchSti1998,Bie2001}.

Experimentally accessible physical systems are usually not closed
and in many cases have a corresponding classical description in which 
trajectories escape or are absorbed.  Such open systems appear, for example,
in optical microcavities \cite{CaoWie2015}, nuclear reactions 
\cite{MitRicWei2010}, or microwave resonators \cite{Sto1999}.
Classically, open systems are characterized by the coexistence of
trapped and (transiently chaotic) escaping motion on the phase-space 
\cite{LaiTel2011,AltPorTel2013}.
Quantum mechanically, one studies complex resonances using a quantum-scattering 
description~\cite{Gas2014b}.
Chaotic resonance eigenfunctions
show fractal distributions \cite{CasMasShe1999b}
and are not universally
described by a single classical measure \cite{KeaNovPraSie2006,NonRub2007}.
%

Best understood are resonance eigenfunctions in open
systems with full escape
\cite{BorGuaShe1991,
    CasMasShe1999b,
    SchTwo2004,
    KeaNovPraSie2006,
    NonRub2007,
    KeaNonNovSie2008,
    ErmCarSar2009,
    ClaKoeBaeKet2018,
    BilGarGeoGir2019},
in which particles are completely absorbed or escape to infinity. Examples are 
scattering systems (e.g., three-discs~\cite{GasRic1989a}) and systems in which 
a phase-space region $\Omega \subset\PS$ acts as a leak~\cite{AltPorTel2013}.
Classically, a fractal chaotic saddle
emerges as
the set of points which never escape for times $t \rightarrow \pm \infty$. 
Typical (smooth) initial distributions converge towards a measure $\munat$, the 
so-called natural
measure, which is conditionally invariant (c-measure), decays with the characteristic natural decay
rate $\gnat$, and is smooth on the
unstable manifold of the saddle \cite{PiaYor1979,DemYou2006}.
Quantum mechanically, resonances are distributed according to a fractal Weyl law
which is determined by the fractal dimension of the chaotic saddle
\cite{Sjo1990, 
    Lin2002, 
    LuSriZwo2003,
    SchTwo2004,
    RamPraBorFar2009, 
    EbeMaiWun2010, 
    ErmShe2010, 
    NonSjoZwo2014}.
The phase-space structure of resonance eigenfunctions depends on their
decay rate  $\gamma$ \cite{KeaNovPraSie2006} and converges in the semiclassical limit to a c-measure~\cite{NonRub2007}.
For $\gnat$ this limit is assumed to be the natural measure $\munat$
\cite{CasMasShe1999b}. For arbitrary decay rates a good description is given
by conditionally invariant measures which are uniformly distributed
on sets with the same temporal distance to
the chaotic saddle \cite{ClaKoeBaeKet2018}.

Less understood are resonance eigenfunctions in
open systems with partial escape,
in which particles escape with some finite probability or their intensities are partially absorbed.
These systems are used in the description of
optical microcavities
\cite{LeeRimRyuKwoChoKim2004,  
    WieHen2008,WieMai2008, 
    ShiWieCao2011,
    ShiHarFukHenSasNar2010, AltPorTel2013b, HarShi2015, CaoWie2015, AltPorTel2015, KulWie2016},
where the
intensity of light rays changes according to Fresnel's law of reflection and
transmission.
Classically, such systems are again described by a natural c-measure $\munat$, which
is multifractal and supported on the full phase-space~\cite{AltPorTel2013b}.
Quantum mechanically, the resonances have finite decay rates,
do not follow a simple fractal Weyl law
\cite{WieMai2008,NonSch2008,PotWeiBarKuhStoZwo2012,
    GutOsi2015,SchAlt2015,KoeMicBaeKet2013},
and the longest living resonances in fully chaotic
optical microcavities are well described by $\munat$ (also known as steady probability distribution) \cite{LeeRimRyuKwoChoKim2004, AltPorTel2013,HarShi2015,KulWie2016}.
Further results were obtained for
the relation to periodic orbits
\cite{CarBenBor2016,PraCarBenBor2018} and
the case of single channel openings
\cite{LipRyuLeeKim2012,LipRyuKim2015}.
However, there is no understanding of how the properties of resonance eigenfunctions depend on their decay rate.
%

In this paper we study the classical origin of the multifractal
phase-space structure of resonance eigenfunctions in systems
with partial escape.
Long-lived eigenfunctions stretch along the unstable direction
of the classical dynamics and are known to
semiclassically correspond to the natural measure.
Short-lived eigenfunctions stretch
along the stable direction and
we conjecture that they semiclassically correspond to
the natural measure of the inverse classical dynamics.
For increasing decay rates the main contribution
changes continuously from unstable to stable direction.
Combining both natural measures we introduce a family
of c-measures which
are uniform on sets with the same decay under
forward and backward time evolution.
Qualitatively, we find that these measures and resonance eigenfunctions
have the same characteristic 
filamentary patterns on phase space
and show a similar dependence on the decay rate.
Quantitatively, we numerically support the conjecture for long- and short-lived 
eigenfunctions.

The paper is divided as follows. In Section~\ref{SEC:QUANTUM} we numerically 
show the emergence of fractal phase-space distributions for
resonance eigenfunctions in an exemplary system.
In Sec.~\ref{SEC:ClassicalMeasures} we show how to construct a family of 
c-measures for all relevant decay rates.
We then compare the resonance eigenfunctions with these
measures in Sec.~\ref{SEC:Comparison}, qualitatively and quantitatively using 
information dimension and Jensen-Shannon divergence. 
A summary is given in Sec.~\ref{SEC:Conclusion}.

\begin{figure}[t!]
    \includegraphics[scale=1.]{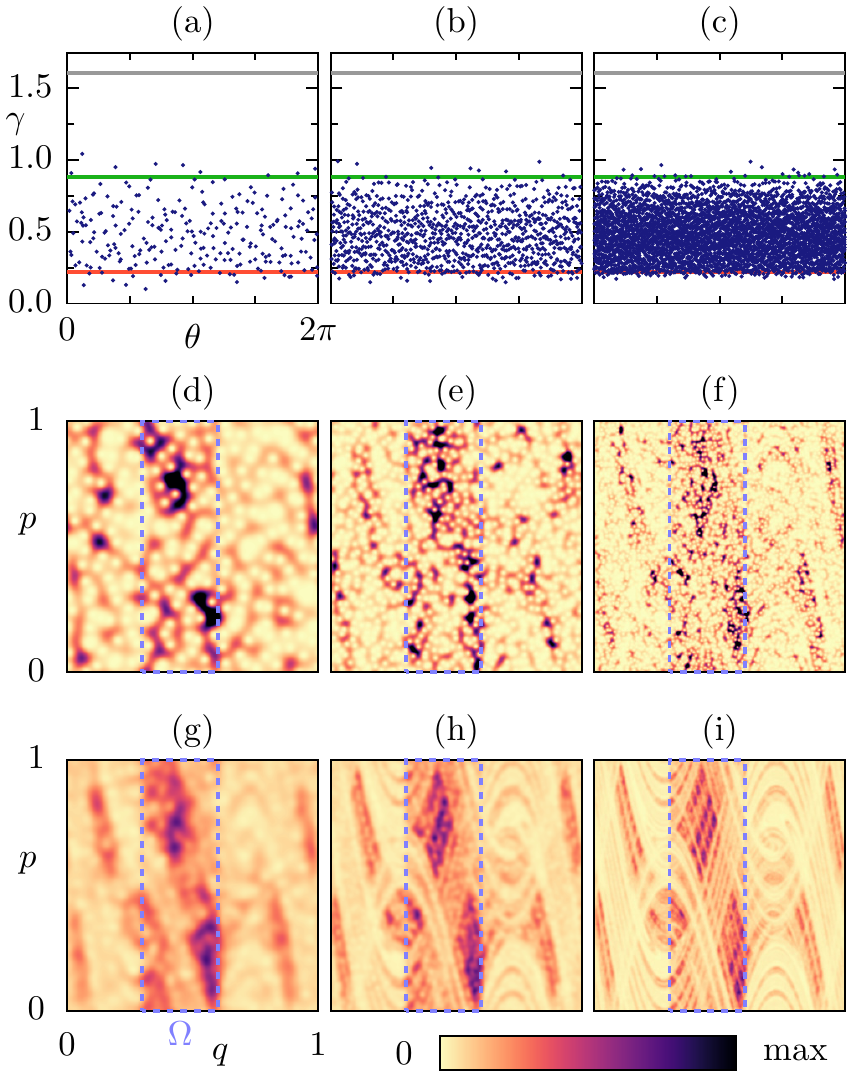}
    \caption{Eigenvalues and fractal eigenfunctions in a chaotic
        quantum map with partial escape. 
        (a-c) Spectrum in $\gamma-\theta$ plane for (a) $h = 1 / 250$,
        (b) $h = 1/1\,000$ and (c) $h=1/4\,000$.
        Horizontal lines indicate classical decay rates
        $\gamma \in \{\gnat, \ginv, \gmax \}$
        (from lower to upper: red, green, gray),  
        defined in Sec.~\ref{SEC:ClassicalMeasures}.
        (d-f) Husimi phase-space distributions $\hus_\gamma$
        of single eigenfunction
        with decay rate closest to $\gamma = 0.55$ for values of $h$ as
        in (a-c). 
        (g-i) Average Husimi distributions
        $\langle\hus\rangle_\gamma$ of eigenfunctions with
        decay rate in $(\gamma - \Delta\gamma, \gamma + \Delta\gamma)$
        with $\gamma = 0.55$ and $\Delta\gamma = 0.016$ for values of $h$ as
        in (a-c),
        using $\{17, 64, 244\}$ eigenfunctions.
        Individual colormap for each $h$ with maximum given by
        $1.3\, \max \langle\hus\rangle_\gamma$.
        Clear emergence of fractal structure.
        The system is the standard map, see App.~\ref{appendix.standardmap},
        on the torus $(q, p) \in [0,1)\times [0, 1)$ with $\kappa = 10$,
        reflection $\alphaOmega= 0.2$, and opening
        $\Omega = (0.3, 0.6) \times[0, 1)$ indicated by dashed lines in (d-i).
    }
    \label{FIG:fig1}
\end{figure}

\section{Resonances in systems with partial escape}
\label{SEC:QUANTUM}
Scattering systems are described by the relation
of incoming waves to outgoing waves,
which is given by the $S$-matrix \cite{Gas2014b}.
Resonances appear as poles of the $S$-matrix and are experimentally
seen as peaks in the absorption spectrum, whenever a resonance
eigenfunction is excited.
A similar description is found, if the dynamics is reducible
to discrete times \cite{FyoSom2000},
e.g.\ in systems with time-periodical driving.
In these systems the $S$-matrix takes the form \cite{KeaNovSch2008}
\begin{equation}
S(\omega) = -\Qrefl +
    \Qtrans \frac{1}{\ue^{-\ui\omega} - \QMapcls\Qrefl}\QMapcls\Qtrans,
    \label{EQ:Smatrix}
\end{equation}
with unitary time-evolution $\QMapcls$ of internal states
and quasi-energy $\omega$.
The internal reflection operator $\Qrefl$ and the transmission
operator $\Qtrans$
govern the coupling from incoming to outgoing waves
and satisfy $\Qrefl^\dagger\Qrefl + \Qtrans^\dagger\Qtrans = 1$,
which expresses the conservation of probability.
Choosing an appropriate basis, the internal reflection
$\Qrefl$ becomes diagonal with positive entries. %
E.g., for microresonators internal reflection
is described by Fresnel's law \cite{KeaNovSch2008,KulWie2016}.
It follows from Eq.~\eqref{EQ:Smatrix} that whenever
$\ue^{-\ui\omega}$ is an eigenvalue of the operator
\begin{equation}
\QMap = \QMapcls \Qrefl,
\label{EQ:QMap}
\end{equation}
a resonance condition is fulfilled.
The operator $\QMap$ describes the time-evolution of internal states
composed of the reflection $\Qrefl$ and unitary time evolution $\QMapcls$.
In the following $\QMap$  is called quantum map with escape.
Solving the eigenvalue equation
\begin{equation}
    \QMap \psi = \ue^{-\ui\theta -\gamma/2} \psi
\end{equation}
gives the complex resonances $\omega = \theta - \ui \gamma/2$
and the (right) resonance eigenfunctions $\psi$.
Note that in time-independent scattering systems 
$\theta$ corresponds to the energy and $\gamma$ to the width
of the resonance.
Applying the quantum map with escape \eqref{EQ:QMap}
$n$-times to an eigenfunction $\psi$
the probability decays like
$\Vert \QMap^n \psi \Vert ^2 = \ue^{-n\gamma}\Vert\psi\Vert^2$.
Therefore $\gamma$ is called \emph{decay rate} of the resonance
eigenfunction $\psi$.
Note that the quantization divides the compact phase space into an integer 
number of $1/h$ Planck cells, where $h$ is the effective Planck's constant.

Figure~\ref{FIG:fig1} shows resonances and resonance eigenfunctions
for a paradigmatic example of
a chaotic quantum system with partial escape (the standard map, see 
Appendix~\ref{appendix.standardmap} for details).
There are two important observations:

\begin{itemize}

\item The spectra show that the resonances have finite decay rates $\gamma$, 
that are mostly inside a relatively narrow band, Fig.~\ref{FIG:fig1}(a-c).
The existence of { spectral gaps}
\cite{NonSch2008,She2008, GutOsi2015}  
explains that the %
decay rates are away from zero and infinity.
In particular, %
there exist
decay rates $\gnat$ and $\ginv$ (explained in the next section) such that 
approximately $0 < \gnat \lesssim \gamma \lesssim \ginv < \gmax$,
where $\gmax$ is determined by the maximal allowed classical escape,
Eq.~\eqref{EQ:DecayLimits} below.

\item The main observation concerns the resonance eigenfunctions
and will be explored and explained in greater generality in the remaining
of the paper.
Resonance eigenfunctions have a rich multifractal
phase-space structure, Fig.~\ref{FIG:fig1}(d-i). 
This structure is in contrast to the uniform distribution observed in closed 
chaotic systems. It depends strongly on the decay rate $\gamma$ (as shown 
below in Fig.~\ref{FIG:fig3}) and is increasingly resolved as $h\rightarrow 0$. 
The fractality 
becomes evident in the Husimi distribution~\cite{Hus1940} obtained averaging 
resonance eigenfunctions with similar decay rate, Fig.~\ref{FIG:fig1}(g-i). 
Single eigenfunctions, Fig.~\ref{FIG:fig1}(d-f), with the same decay rate 
$\gamma$ have similar
phase-space distributions up to fluctuations around the average. Fluctuations 
around the average are expected to originate from quantum properties, like in 
closed systems~\cite{NonVor1998}.
  
\end{itemize}

In the next section we search for classical explanations for the observations summarized above. In particular, the convergence of resonance eigenfunctions to fractal distributions motivate us to search for corresponding measures of the classical system.

\section{Classical c-measures}
\label{SEC:ClassicalMeasures}

In this section we construct classical measures with
decay rates $\gamma$ that act as candidates to describe the resonance eigenfunctions with the same decay rate $\gamma$ for $h \to 0$.
We define the classical analogue of the quantum map with escape as a time discrete chaotic map
$\Mapcls: \PS \rightarrow \PS$ on a bounded phase space $\PS$, and
a classical reflection function
$\refl:\PS\rightarrow\mathbb{R}_+$.
We assume $\Map$ to be measure preserving with respect
to the Lebesgue measure $\muL$.
The classical map with escape is the application of
the reflectivity $\refl$ on the intensity followed by the closed
time-evolution $\Mapcls$.
This is written in the extended phase-space $\PS \times \mathbb{R}_+$
as the mapping of initial points $x\in \PS$ with intensity $J\in \mathbb{R}_+$, i.e.\
$(x', J') = (\Mapcls(x), J \cdot \refl(x))$.

The time evolution of phase-space points with an assigned intensity
is generalized to measures in the following sense.
For a given map $M$ and reflection function $R$ we define for
any measure $\mu$ on $\PS$ the \textit{classical map with escape}
$\MMap\equiv \MMap_{\Map,\refl}$ as
\begin{equation}
    \MMap\mu(A) := \int_{\Map^{-1}(A)}R\,\ud\mu,  \label{EQ:MMap0}
\end{equation}
for all measurable $A\subset\PS$. %
Thus the weight of the set $A$ for the iterated measure $\MMap\mu$
is given by the measure of the preimage $\Mapcls^{-1}(A)$ weighted with
the reflectivity $\refl(x)$.
This definition of $\MMap$ ensures that
for a single phase-space point $x\in \PS$ the intensity is
first reduced by $\refl(x)$, followed by iteration with $\Mapcls$.

A measure $\mu$ is called conditionally invariant (c-measure) %
with respect to the map with escape $\MMap$, if
\begin{equation}
    \MMap\mu(A) = \ue^{-\gamma}\,\mu(A)   \label{EQ:cinv}
\end{equation}
for all measurable $A\subset\PS$
with eigenvalue $\ue^{-\gamma} \in \mathbb{R}_+$
\cite{PiaYor1979,DemYou2006}.
Here $\gamma$ is the decay rate of the
c-measure, as for $n$ iterations
$\Vert\MMap^n\mu\Vert = \ue^{-n\gamma} \Vert\mu\Vert$ with norm
$\Vert \mu \Vert := \mu(\PS)$.
Possible values of decay rates $\gamma$ are bounded by
minimal  and maximal values of the reflectivity as \cite{NonSch2008}
\begin{align}
 - \log (\max_{\Gamma} \refl) \equiv \gmin\leq \gamma
\leq \gmax \equiv -\log (\min_{\Gamma} \refl).
\label{EQ:DecayLimits}
\end{align}

Resonance eigenfunctions with quantum decay rate $\gamma$
are expected to
converge in the semiclassical limit to c-measures with the corresponding
classical decay rate, as in systems with full escape \cite{NonRub2007}.
For maps with partial escape, as considered here, the importance of 
multifractals has been emphasized for the natural decay
(corresponding to the largest eigenvalue of 
the Perron-Frobenius operator)~\cite{AltPorTel2013b,SchAlt2015}.
Here we investigate fractal properties for different decay rates $\gamma$
by  constructing c-measures of the classical dynamics.
We start with the construction for the 
natural measures and use them to obtain c-measures in the full range 
$\gmin < \gamma < \gmax$. 

\subsection{Natural measure}

If the dynamics on phase-space is ergodic and hyperbolic,
smooth initial distributions asymptotically decay
with one characteristic rate $\gnat$ and approach
the corresponding {natural c-measure} $\munat$
\cite{CheMar1997a,CheMarTro2000,DemYou2006}.
This can be used to construct $\munat$ by time-evolution
of the uniform Lebesgue measure $\muL$
and normalization, %
\begin{align}
\munat(A) &= \lim\limits_{n\rightarrow\infty}
\frac{\MMap^n\muL(A)}{\Vert \MMap^n\muL\Vert}, \quad \text{with}
\label{EQ:munatlim} \\
\MMap^n\muL(A) &= \int_A \prod_{i=1}^{n}\refl[\Map^{-i}(x)]\,\ud x\text{.}
\label{EQ:munat}
\end{align}
Equation~\eqref{EQ:munat} follows from successive application of
Eq.~\eqref{EQ:MMap0} and integral transformation with measure preserving 
$\Mapcls$.
It has the following intuitive interpretation:
Phase-space points which experience the same average decay
under $n$ backward iterations get the same weight. 
Thus $\munat$ is uniformly distributed on sets with the same average
decay under backward iteration.

\subsection{Natural measure of the inverse map}
\label{SEC:Inverse}
In the following we use the inverse map to identify another
important classical c-measure and its decay rate.
The classical map with escape $\MMap$ is invertible,
if the reflectivity $\refl(x) > 0$ for almost all $x\in\PS$.
This is for example
the case for TE and TM polarization in optical microcavities.
The inverse $\MMap^{-1}$ is given by application of the inverse map
$\Map^{-1}$ on $\Gamma$
followed by the inverse reflectivity $\refl^{-1}\equiv 
1/\refl$~\cite{AltPorTel2015}.
Note the different meanings of the exponent $-1$ for both $R$ and $M$.
Thus for measures $\mu$ on $\PS$ we obtain the inverse map
\begin{equation}
    \MMap^{-1} \mu(A) =  \int_{\Map(A)}\refl^{-1}\circ\Map^{-1}\,\ud\mu
\end{equation}
which has the same form as Eq.~\eqref{EQ:MMap0}
and satisfies $\MMap\circ\MMap^{-1} = \MMap^{-1}\circ\MMap = \mathds{1}$.
The inverse map itself is again a chaotic map with escape and is given by
$\MMap^{-1} = \MMap_{\Map^{-1}, \refl^{-1}\circ\Map^{-1}}$.
Therefore, results obtained for $\MMap$ are similarly valid for
$\MMap^{-1}$.
We first conclude that c-measures $\mu$ of $\MMap^{-1}$ with decay rate
$\gamma$ are also 
c-measures of $\MMap$ with decay rate $-\gamma$ and vice versa.
Secondly, there exists a natural measure
of the inverse map, which we will call \textit{inverse measure} of $\MMap$,
\begin{align}
\muinv &:= \munat[\MMap^{-1}], \qquad \text{with} \\
\ginv &= -\gnat[\MMap^{-1}].
\end{align} 
Note that $\ginv$ and $\gnat$ are independent 
\cite{AltPorTel2015} and $\ginv > \gnat$.
Note also that the inverse decay rate $\ginv$ is not related
to the so-called inner edge observed in the Walsh-quantized
Baker map with full escape \cite{KeaNonNovSie2008}.

There are several possibilities to obtain $\muinv$, e.g.,
    using the Ulam method \cite{Ula1960,ErmShe2010}
    for the Perron-Frobenius operator
    of the inverse map $\MMap^{-1}$.
The construction using time-evolution
similar to Eq.~\eqref{EQ:munatlim} is given by
\begin{align}
    \muinv(A) &= \lim\limits_{m\rightarrow\infty}
    \frac{\MMap^{-m}\muL(A)}{\Vert\MMap^{-m}\muL\Vert}, \quad\text{with}
    \label{EQ:muinvlim}\\
\MMap^{-m}\muL(A) &= \int_A \prod_{i=0}^{m-1}\refl^{-1}[\Map^i(x)]\,\ud 
    x\text{.} \label{EQ:muinv}
\end{align}
In contrast to $\munat$
the inverse measure $\muinv$ is uniformly distributed on sets with the
same average decay under forward iteration.

\begin{figure}[t!]
    \includegraphics[scale=1.]{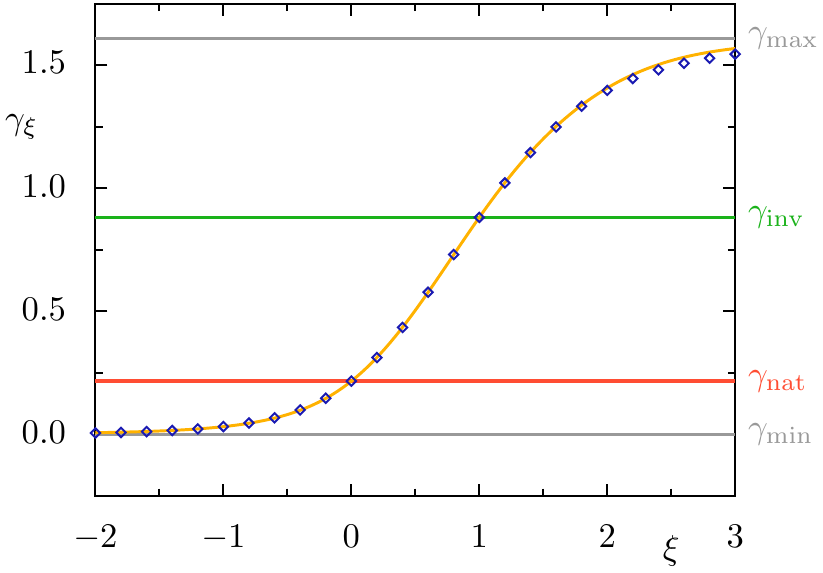}
    \caption{%
    Classical decay rates $\gamma_\xi$ vs.\  $\xi$.
    Numerical evaluation of Eq.~\eqref{EQ:GammaXi} (orange) from decay
    rates of corresponding natural and inverse measures.
    Decay rate $\gamma_\xi = -\log \MMap\muxi(\Gamma)$ based on 
    Eq.~\eqref{EQ:cinv} for numerically constructed $\muxi$ (blue diamonds).
    Horizontal lines indicate classical decay rates
    $\gamma \in \{\gmin, \gnat, \ginv, \gmax \}$.
    Standard map as in Fig.~\ref{FIG:fig1}.
        }
    \label{FIG:fig2}
\end{figure}

\begin{figure*}[t!]
    \includegraphics[scale=1.]{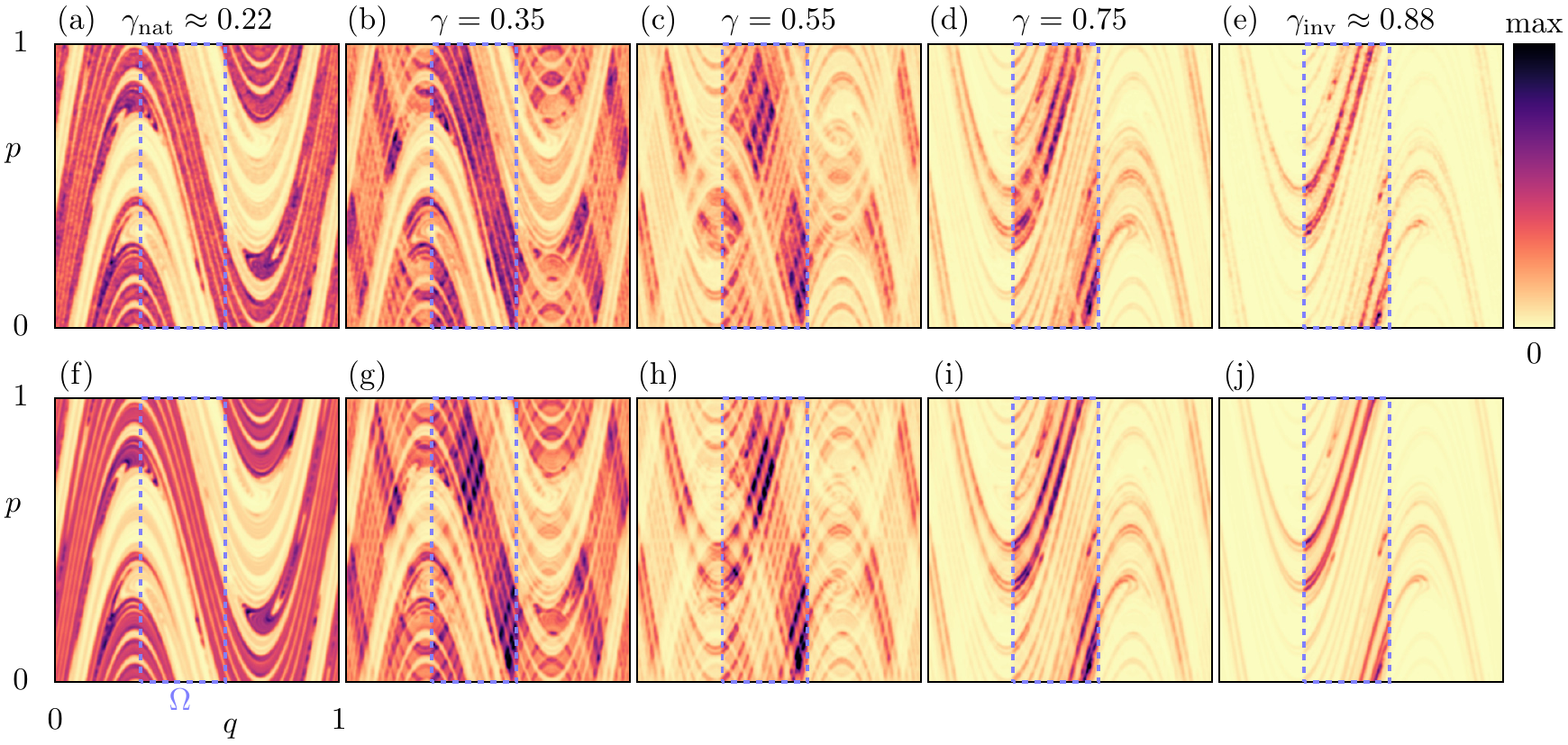}
    \caption{%
        Quantum (top) to classical (bottom) correspondence.
        (a-e) Average Husimi distributions
        $\langle\hus\rangle_\gamma$ of resonance eigenfunctions with
        $\gamma \in \{\gnat\approx0.22, 0.35, 0.55, 0.75, \ginv\approx0.88\}$
        and $\Delta\gamma = 0.016$ for $h = 1/4\,000$,
        using $\{152, 248, 244, 127, 20\}$ eigenfunctions.
        Individual colormap for each $\gamma$ with maximum given by
        $\max \langle\hus\rangle_\gamma$.
        (f-j) Gaussian smoothed phase-space distribution of
        c-measures $\muxi$ for
        $\xi \in \{0, 0.266, 0.563, 0.825, 1\}$ corresponding to the
         same
        $\gamma$ as in (a-e) using the same colormaps, respectively.
        The opening $\Omega$  is indicated by dashed lines.
        Standard map as in Fig.~\ref{FIG:fig1}.
    }
    \label{FIG:fig3}
\end{figure*}

\subsection{C-measures for arbitrary \ensuremath{\gamma}}
\label{SEC:Construction}
In the following we use the natural and the inverse measure to construct
c-measures with arbitrary decay rate $\gamma$.
The main idea is to use the local phase-space structure of stable and unstable
directions in hyperbolic maps.
While $\munat$ is smooth (fractal) along the unstable (stable)
direction, $\muinv$ is smooth (fractal) along the stable (unstable)
direction.
The fractal distribution is responsible for fulfilling conditional
invariance, i.e.\ the partial escape with $\refl$ and iteration
with $\Mapcls$ leads to the global decay factor
$\ue^{-\gnat}$ and $\ue^{-\ginv}$, respectively.
Factorizing the reflectivity 
$\refl = \refl^{1 - \xi}\refl^\xi$ for $\xi \in \mathbb{R}$
it is possible to consider the local product of
the natural measure for reflectivity $\refl^{1-\xi}$, $\munat[\refl^{1-\xi}]$, 
and the inverse measure for reflectivity $\refl^\xi$, $\muinv[\refl^\xi]$.
This gives a c-measure for the reflectivity $\refl$ with decay rate
\begin{equation}
\gxi = \gnat[R^{1 - \xi}] + \ginv[R^\xi], \label{EQ:GammaXi}
\end{equation}
see App.~\ref{appendix:proof-ci} for additional motivation.
The explicit construction is given by
\begin{align}
    \muxi(A) &= \lim\limits_{n\rightarrow\infty} 
    \frac{\tilmuxi^n(A)}{\Vert\tilmuxi^n\Vert}, \quad \text{with}\\
    \tilmuxi^n(A) &= \int\limits_A
    \prod_{i=1}^{n}\refl^{1 - \xi}[\Map^{-i}(x)]\,
    \prod_{j=0}^{n-1}\refl^{-\xi}[\Map^j(x)]\,\ud x.
    \label{EQ:muxi}
\end{align}
This measure is uniformly distributed on sets 
with the same average decay under backward iteration and  the same average
decay under forward iteration.

For $\xi = 0$ just the first factor contributes in Eq.~\eqref{EQ:muxi}
and we recover $\mu_{\xi = 0} = \munat$. From Eq.~\eqref{EQ:GammaXi} follows
that $\gamma_{\xi = 0} = \gnat[\refl] = \gnat$,
where we used that $\ginv[\refl^0] = 0$.
For $\xi = 1$ we similarly have $\mu_{\xi = 1} = \muinv$ and
$\gamma_{\xi = 1} = \ginv$.
Increasing $\xi$ from $0$ to $1$ a transition from $\munat$ to 
$\muinv$ occurs. In Fig.~\ref{FIG:fig2} we show the dependence of the decay 
rate $\gamma_\xi$ on $\xi$,  
Eq.~\eqref{EQ:GammaXi},
for the example system.
The decay rate $\gxi$ continuously increases with $\xi$ from
$\gmin$ for $\xi \rightarrow -\infty$ to $\gmax$ for $\xi \rightarrow \infty$.
Furthermore we determine $\gxi$ by iteration of
the constructed measures $\muxi$ using Eq.~\eqref{EQ:cinv} 
with very good agreement with Eq.~\eqref{EQ:GammaXi}.
The phase-space structure of $\muxi$ is shown in Fig.~\ref{FIG:fig3}(f-j),
demonstrating the strong dependence on the
    decay rate and revealing the underlying hyperbolic structure.
Note that for a closed map, i.e.\ without escape, $\refl(x) = 1$,
we obtain $\gxi=0$ and uniform distribution $\muxi = \muL$ for all $\xi$.

\section{Quantum-classical comparison}
\label{SEC:Comparison}

In this section we investigate to which extent the phase-space 
distributions of resonance
eigenfunctions with decay rate $\gamma$ are described
by the c-measures $\muxi$ with $\gxi = \gamma$.
This comparison is essential because for the same decay rate $\gamma$ there may be many different
c-measures \cite{DemYou2006,NonRub2007} and it is not known which of the possible
c-measures are quantum mechanically relevant in the case of partial escape.
It is known that resonance eigenfunctions with decay rate $\gamma \approx \gnat$
are well described by the natural measure $\munat$ \cite{CasMasShe1999b}, also
in optical microcavities showing partial escape~\cite{LeeRimRyuKwoChoKim2004,HarShi2015,KulWie2016}.
Our first conjecture is that resonance eigenfunctions with decay rate
$\gamma \approx \ginv$
are well described by the inverse measure $\muinv$ in the semiclassical limit.
This is motivated by using that the natural measure of the inverse map
$\MMap^{-1}$ describes resonance eigenfunctions with
$\gamma \approx \gnat[\MMap^{-1}]$ of the inverse
quantum map $\QMap^{-1} = \Qrefl^{-1}\QMapcls^\dagger$.
These are the resonance eigenfunctions of $\QMap$ with decay rate
$\gamma \approx \ginv$.
Our second conjecture is that resonance eigenfunctions with arbitrary decay rate $\gamma$
are well described by the c-measures $\muxi$ with $\gxi = \gamma$.

The quantum-classical comparison is done by qualitatively comparing the 
phase-space distributions
(Sec.~\ref{SEC:Qualitative}),
comparing the information dimension (Sec.~\ref{SEC:Information}),
determining the Jensen-Shannon divergence 
(Sec.~\ref{SEC:QuantitativeComparison}),
studying the semiclassical limit (Sec.~\ref{SEC:Semiclassical}),
and considering the limit of full escape (Sec.~\ref{SEC:FullEscape}).
We report numerical 
investigations for a representative example of fully chaotic system with 
partial 
absorption, the quantized standard map (see 
Appendix~\ref{appendix.standardmap}).
We find very good agreement for the natural measure $\munat$
and the inverse measure $\muinv$ (first conjecture),
and identify small deviations
in the semiclassical limit for intermediate decay rates $\gxi$ (second conjecture).

\subsection{Phase-space distributions}
\label{SEC:Qualitative}

Figure~Fig.~\ref{FIG:fig3} shows a remarkable similarity between the quantum 
Husimi 
distributions and the classical c-measures $\muxi$ for different values of $\gamma$. 
The same multifractal structure and dramatic change with $\gamma$ are observed: 
For $\gamma 
\approx \gnat$ the high-density filaments are concentrated along the
unstable direction on the phase space and resemble the natural c-measure
$\munat$.
For $\gamma \approx \ginv$ the distribution concentrates along
the stable direction %
with maximum inside the opening $\Omega$,
resembling the inverse measure $\muinv$.
For intermediate values of $\gamma$ 
the density concentrates on the product of the two previous 
structures, revealing the hyperbolic structure on phase space.
Classically this behavior is understood from  the definition of $\muxi$,
see Sec.~\ref{SEC:Construction}.

Note that the measures $\muxi$ are illustrated on phase space by considering
expectation values of Gaussian distributions $g_{x,\sigma}$,
centered at $x\in\PS$ with width $\sigma = \sqrt{\hbar/2}$.
This is the classical equivalent to the quantum expectation
of an eigenfunction $\psi$ in a symmetric
coherent state centered at $x\in\PS$.
It is possible to choose asymmetric coherent states
with different uncertainties in the phase-space variables
$q$ and $p$ in the definition of the Husimi distribution.
We find good qualitative agreement, if
the classical expectation values are adapted accordingly (not shown).

The qualitative similarity of quantum and classical distributions confirms that 
the main features of the $\gamma$-dependence of resonance eigenfunctions
has a classical origin. However, a closer inspection shows that there are small 
but still visible differences between the classical and quantum results (e.g.\ 
for $\gamma = 0.55$ in the region $(q, p)\in [0.3, 0.4]\times[0.3, 0.6]$, see 
Fig.~\ref{FIG:fig3}(c, h). This motivates us to pursue more quantitative 
comparisons between the measures, which will allow us to investigate their 
dependence on $\gamma$ and $h$.

\begin{figure}[b!]
    \includegraphics[scale=1.]{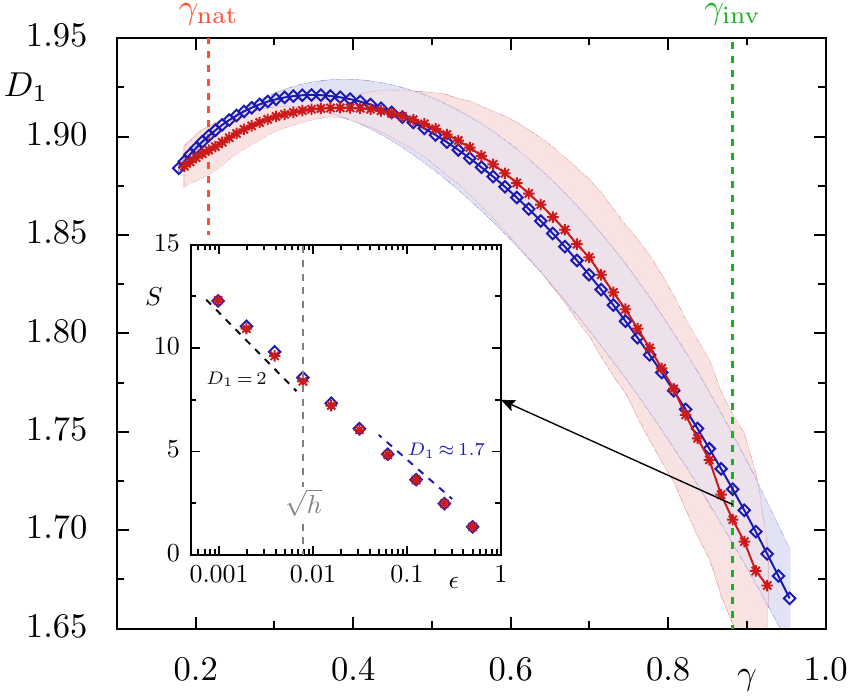}
    \caption{Fractality of classical and quantum measures. The effective information dimension
        $\Dim $ is shown as a function of the decay rate $\gamma$
        for c-measures $\mu_{\xi}$ (blue diamonds)
        and for Husimi distributions
        $\hus_\gamma$ at $h = 1/16\,000$ 
        (red stars).
        $\Dim$ is computed from the scaling
        $S(\epsilon ) \sim \log \epsilon^{-\Dim}$ 
        in the regime $\epsilon \in [1/4, 1/16]$, see inset for the case 
        $\gamma = \ginv \approx 0.88$ ($\Dim = 2$  is expected for the quantum 
        data in the regime $\epsilon \lesssim \sqrt{h}$, black dashed line).
        In the quantum case, $S(\epsilon)$ corresponds to the average entropy 
        for
        eigenfunctions with decay rate in
        $(\gamma - \Delta\gamma, \gamma + \Delta\gamma)$ with
        $\Delta\gamma = 0.016$
        (equivalent results are obtained using the entropy of
            the average Husimi distribution).
        The shaded regions are confidence intervals estimated taking into 
        account uncertainties in the linear regression and (for the quantum 
        case) in the computation of average $S(\epsilon)$.
        Vertical dashed lines indicate $\gnat$ and $\ginv$.
        Standard map as in Fig.~\ref{FIG:fig1}.
    }
    \label{FIG:fig4}
\end{figure}

\subsection{Information dimension}\label{SEC:Information}

In order to quantify the fractal properties of quantum and classical
phase-space distributions
we consider %
the information dimension $\Dim$. It is defined for 
any measure $\mu$ as
\begin{equation}
\Dim[1] = \lim_{\epsilon\rightarrow 0} \frac{S(\epsilon;\mu)}{\log 
    \epsilon^{-1}},
\end{equation}
where the entropy $S(\epsilon; \mu)$ is
defined for any normalized phase-space measure $\mu$ as
\begin{equation}
S(\epsilon;\mu) = -\sum_{A\in\mathcal{A}_\epsilon} p_A \log p_A,
\label{EQ:Entropy}
\end{equation}
with $    p_A \equiv \mu(A) = \int_A \ud\mu$ and $\mathcal{A}_\epsilon$ a
partition of the phase-space in boxes $A$ of size $\muL(A) = \epsilon^2$.
We choose $\Dim[1]$ instead of $\Dim[0]$
to quantify the fractality of the measures because $\hus(x) 
> 0$ almost everywhere on $\PS$ and therefore the box-counting dimension 
$\Dim[0]$ is equal to the phase space dimension~\cite{AltPorTel2013b}.
For the 
Husimi distribution $\hus$ the probability measure is
defined as $\mu_\hus(A) = \int_A \hus\, \ud\muL$ (see 
Appendix~\ref{appendix.entropy} for details).
As it is a smooth function on the scale of order $h$,
any asymptotically ($\epsilon \rightarrow 0$) defined fractal dimension of $\hus$ is trivial.
 Therefore, we focus on an effective fractal dimension $D_1$ in
 a regime $1 > \epsilon \gtrsim \sqrt{h}$.
 The dimension $\Dim[1]$ (and entropies $S$) for resonance eigenfunctions with 
 the same decay rate $\gamma$ should converge
in the limit $h \rightarrow 0$ for a fixed value of $\epsilon$.

Figure~\ref{FIG:fig4} shows numerical results for $\Dim[1]$ of the quantum and 
classical measures. The inset confirms that $S$ shows a non-trivial scaling 
with $\epsilon$, defining an effective fractal dimension for the
quantum (red stars) and classical (blue diamonds)
distributions.  The dependence of $\Dim $ on the decay 
rate $\gamma$ shows an initial increase to a maximum value, followed by a 
decrease towards $\Dim(\ginv) < \Dim (\gnat)$.
This shows that different fractal properties exist for different decay rates 
$\gamma$. Most importantly, the quantum and classical results show the same 
overall dependence.
The distance between the values for $\muxi$ and $\hus_\gamma$ are smaller than 
the numerically estimated errors from the calculation of $\Dim $.
Therefore the fractal dimension of the c-measures $\muxi$
can be used to estimate the fractal dimension of resonance eigenfunctions.

\begin{figure}[b!]
    \includegraphics[scale=1.]{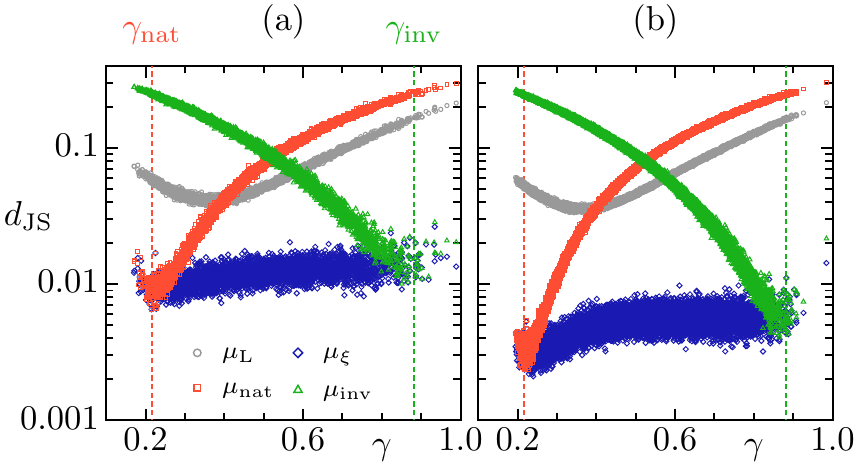}
    \caption{Comparison between the quantum measure and different classical 
    measures as a function of the decay rate $\gamma$. The symbols correspond 
    to the Jensen-Shannon divergence $\jsd$ between individual 
    Husimi distributions $\hus_\gamma$ and different classical measures
        $\muxi$ (blue diamonds),
        $\munat$ (red boxes),
        $\muinv$ (green triangles),
        and $\muL$ (gray circles). (a) $h = 1/4\,000$ and (b) $h=1/16\,000$.
        Dotted lines indicate  $\gnat$ and $\ginv$.
          The reported distances $\jsd$ were obtained with $\epsilon=1/16 > 
          \sqrt{h}$.
        Standard map as in Fig.~\ref{FIG:fig1}.
        }
    \label{FIG:fig5}
\end{figure}
\subsection{Jensen-Shannon divergence}
\label{SEC:QuantitativeComparison}

A quantification of the distance between the quantum and classical measures can 
be obtained using the Jensen-Shannon divergence~\cite{GroBerCarRomOliSta2002}.
For any two probability measures $\mu_1$ and $\mu_2$ it is given by the
difference of the entropy of the sum of the distributions and the
sum of the entropies of the single distributions,
\begin{align}
\jsd (\epsilon; \mu_1, \mu_2) = S\left(\epsilon; \frac{\mu_1 + 
\mu_2}{2}\right)
       - \frac{S(\epsilon; \mu_1) + S(\epsilon; \mu_2)}{2},
\end{align}%
where $S$ is the entropy defined in Eq.~\eqref{EQ:Entropy}.
The square root of $\jsd$ is a distance and thus $\jsd(\mu_1, \mu_2) \geq 0$ with equality
if and only if $\mu_1 = \mu_2$. Again we need to introduce
a scale of order $\epsilon$ defining the coarseness of the phase-space.
For quantum-classical comparison
the calculated difference is only meaningful if $\epsilon > \sqrt{h}$.

The Jensen-Shannon divergence between individual
Husimi distributions and
several classical measures is illustrated in Fig.~\ref{FIG:fig5}(a)
for different values of $\gamma$ and $h$. %
As expected, the distance to the natural measure $\munat$ (red boxes) has a 
pronounced minimum at $\gamma = \gnat$ and the distance to $\muinv$ at $\ginv$ 
(green triangles).
The distance to the uniform measure $\muL$ (gray circles) is larger than these 
minima, showing a minimum  at around $\gamma \approx 0.4$ (related to the 
maximum of $\Dim $ seen in Fig.~\ref{FIG:fig4},
where $\hus_\gamma$  is closer to uniformity).
The distance to the c-measures $\muxi$ (blue diamonds) shows
much smaller values of $\jsd$ than for the other measures.
There is almost no dependence on $\gamma$.
Reducing $h$ in Fig.~\ref{FIG:fig5}(b), we see that the quantum-classical 
distance reduces significantly for all $\gamma$.
Variation of $\epsilon > \sqrt{h}$ gives similar results
(with overall smaller distances for larger $\epsilon$ and vice versa). 

\begin{figure}[t!]
    \includegraphics[scale=1.]{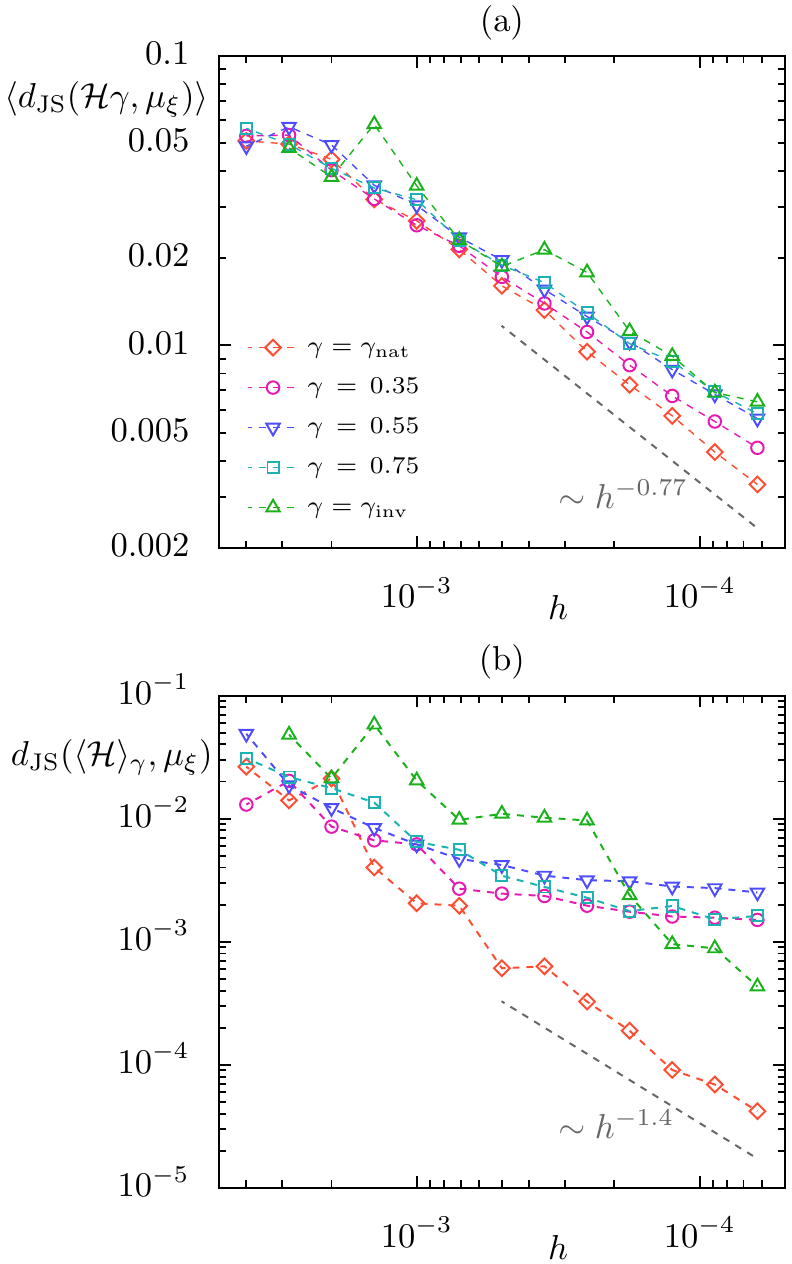}
    \caption{Quantum-classical comparison in the semiclassical limit. The
        Jensen-Shannon divergence $\jsd$ between quantum
        distributions and the classical measure $\muxi$ is shown as a function of $h$
        for $\gamma \in
        \{ \gnat\approx 0.22, 0.35, 0.55, 0.75, \ginv\approx 0.88\}$.
        (a) Husimi distribution of individual resonances $\hus_\gamma$, 
        averaged over $\Delta\gamma = 0.004$.
        Gray dashed line indicates numerical scaling for $\gnat$.     
        (b) Averaged Husimi distributions $\husavg$
        with $\Delta\gamma = 0.004$.
        Standard map as in Fig.~\ref{FIG:fig1}.
    }
    \label{FIG:fig6}
\end{figure}

\subsection{Semiclassical limit}
\label{SEC:Semiclassical}
We focus on the quantum-classical comparison 
with  the aim of testing whether
our results are compatible with a distance $\jsd \rightarrow 0$
in the semiclassical limit, $h \rightarrow 0$. 
Distances $\langle \jsd(\hus_\gamma, \muxi)\rangle$
for a fixed scale $\epsilon = 1/16 \gg \sqrt{h}$ and five 
different decay rates $\gamma$ are shown in Fig.~\ref{FIG:fig6}(a),
comparing individual Husimi distributions $\hus_\gamma$ and the classical 
measures $\muxi$, averaged over $\Delta\gamma = 0.004$.
We observe a power-law $\jsd \sim h^\delta$ decay with $\delta \approx 0.77$. %
Distances of the same order and scaling are found
    when comparing individual Husimi distributions $\hus_\gamma$ to the
    averaged Husimi distributions $\husavg$ (not shown).
Therefore at these values of $h$ the distance
and its decay is dominated by fluctuations around the average.
Establishing a possible relationship between $\delta$ and
fractal properties of $\muxi$ remains an open question.

In order to obtain %
a more sensitive test we reduce the quantum 
fluctuations by computing averaged Husimi 
distributions $\husavg$ using eigenfunctions in an interval around $\gamma$.
The results presented in Fig.~\ref{FIG:fig6}(b)
show therefore much smaller distances $\jsd$.
For $\gnat$ and $\ginv$ we again find a power-law (with larger exponent),
indicating semiclassical convergence.
This gives strong evidence for our conjecture about the inverse measure
$\muinv$. It also verifies the expectation
\cite{CasMasShe1999b,LeeRimRyuKwoChoKim2004}
for the natural measure $\munat$ on a quantitative level.

For intermediate values of $\gamma$, however,
we observe a much slower decay of  $\jsd$ with $h$.
This suggests a saturation towards a finite distance
$\jsd(\husavg^{h\rightarrow0}, \muxi) > 0$.
Thus the c-measures $\muxi$ are not the semiclassical limit measures
of the resonance eigenfunctions.
We expect a similar saturation also for the individual
Husimi distributions, as used in Fig.~\ref{FIG:fig6}(a). However, this is 
expected for smaller values of the distance $\jsd$ (when it is no longer 
dominated by quantum fluctuations but by the distance of the
classical measure to the average Husimi distribution)
which occur for values of $h$ beyond the 
reach of current computational feasibility.
%

\subsection{Limit of full escape}
\label{SEC:FullEscape}

Finally we investigate the
dependence on the reflectivity $\alphaOmega$. Our construction of $\muxi$ can 
be performed for any system with partial escape ($\alphaOmega \ne 0$) and our 
results hold, with quantitative changes due to the change in the fractality
of the c-measures. For instance, in the limit of a closed system, $\alphaOmega 
\rightarrow 1$, all $\muxi$ approach the uniform (Liouville) measure. The most 
interesting case is the
limit of full escape, i.e.\ when
the reflectivity function only takes values $0$ or $1$.
In our example system this corresponds to $\alphaOmega = 0$
in the opening and $\refl = 1$ in the remaining part of phase space.
Classically, regions of full and partial escape are typically treated 
differently, e.g., in the operator proposed in Ref.~\cite{AltPorTel2013b}. Most 
importantly, when regions of full escape exist, the open map $\MMap$ is not 
invertible, so that $\muinv$ is not well-defined and the measures $\muxi$ 
cannot be constructed directly. In order to investigate how $\muxi$ behaves in 
the case 
of full escape, here we consider the limit $\muxi[\alphaOmega\rightarrow 0]$.
It is important in this limit to adjust the parameter $\xi$ 
such that the decay rate $\gamma$ remains fixed.
In this way we obtain a c-measure with decay rate $\gamma$ for the system
with full escape.

\begin{figure}[t!]
    \includegraphics[scale=1.]{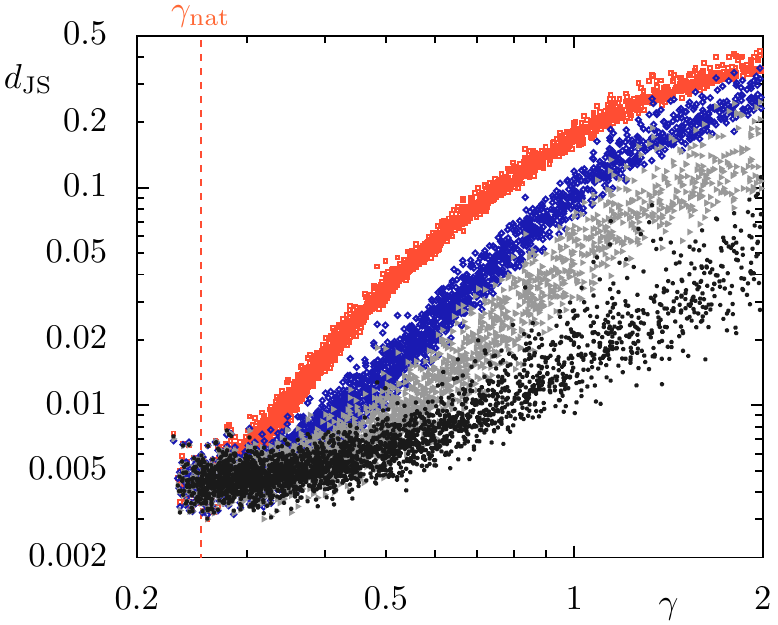}
    \caption{Quantum-classical comparison for systems with full escape. The
        Jensen-Shannon divergence $\jsd(\epsilon=1/16)$ is shown as a function of $\gamma$ for
        fixed $h = 1/16\,000$. The Husimi distribution
        $\hus_\gamma$ of individual eigenfunctions is compared to different classical measures:
        $\munat$ (red squares),
        $\muxi$ for small  $\alphaOmega = 10^{-4}$ 
        (blue diamonds),
        c-measures of Ref.~\cite{ClaKoeBaeKet2018} (black circles),
        and c-measures of Ref.~\cite{KoeBaeKet2015}
        (gray triangles).
        Standard map as in Fig.~\ref{FIG:fig1}, but with $\alphaOmega = 0$.
        The dashed line indicates $\gnat\approx 0.25$ for this case.}
    \label{FIG:fig7}
\end{figure}

In Fig.~\ref{FIG:fig7} we show the Jensen-Shannon divergence $\jsd$
comparing resonance eigenfunctions of the system with full escape
($\alphaOmega = 0$)  %
to the measures $\muxi[\alphaOmega = 10^{-4}]$ (blue diamonds).
We verified that results do not change for smaller $\alphaOmega$.
While we find good agreement for decay rates close to $\gnat$, the
distance grows drastically with $\gamma$.
Even though the distance is smaller than for $\munat$ (red squares),
the measures $\muxi$ clearly do not correspond to
resonance eigenfunctions in this limit.
This discrepancy is not surprising,
since the $\gamma$-dependence of $\muxi$ is partly based on information about
forward iterations of phase-space points, which is completely lost
when falling into the opening with $\alphaOmega = 0$.  

Much smaller distances are obtained using the c-measures proposed for systems
with full escape in Ref.~\cite{ClaKoeBaeKet2018} (black circles).
Their construction is based on a uniform distribution on subsets with the
same temporal distance to the chaotic saddle, which is the invariant set
of the system with full escape.
Using the Jensen-Shannon divergence we are able to quantify their
agreement to resonance eigenfunctions.
We observe that the distance grows with $\gamma$, seen already qualitatively
in Ref.~\cite{ClaKoeBaeKet2018}.
Note that the quantitative analysis confirms that these measures are
in better agreement than the c-measures of Ref.~\cite{KoeBaeKet2015}
(light gray triangles).
Note also that for $\gamma = \gnat$ all considered measures are identical
explaining why they have the same distance $\jsd$.

\section{Conclusions}
\label{SEC:Conclusion}

In summary, we construct a one-parameter family of classical conditionally 
invariant measures (c-measures) $\muxi$ that explains the main properties of 
quantum resonance eigenfunctions in systems with partial escape. We confirm 
previous observations that the natural c-measure $\munat=\mu_{\xi=0}$ describes 
long-lived eigenfunctions with decay rate $\gamma=\gnat$ in the semiclassical 
limit $h\to 0$, including numerical tests of this correspondence with 
unprecedented precision. We find a similar good numerical agreement between 
the natural c-measure of the inverse system $\muinv=\mu_{\xi=1}$ and  
short-lived eigenfunctions with $\gamma = \ginv$,
supporting our conjecture about the inverse measure.

The importance of our results is that they apply to arbitrary decay rates 
$\gamma$.
We numerically observe that the decay rates of almost all resonances
lie in the interval $ \gnat \leq \gamma \leq \ginv$.
Our construction of $\muxi$, Eq.~\eqref{EQ:muxi}, is based on the product 
structure in hyperbolic systems and combines $\munat$ and $\muinv$. It leads to 
measures in the complete range of classical decay rates, i.e.,
$\gamma_{\xi \to -\infty} = \gmin$ and $\gamma_{\xi \to \infty} = \gmax$, being 
thus applicable to describe all possible resonances.
Numerical simulations in the standard map show that $\muxi$ captures the main 
features of the quantum resonance eigenfunctions with $\gamma = \gamma_\xi$: 
They show compatible fractal dimensions and the same drastic dependence of the 
phase-space structure on the decay rate $\gamma$ changing from unstable to 
stable
direction. The semiclassical behavior $h\rightarrow 0$ gives strong evidence that for $\gnat$ and $\ginv$ the corresponding measures $\munat$ and $\muinv$ are the semiclassical
limits of resonance eigenfunctions, while for intermediate rates we find no 
convergence. We also find that in the limit of full escape the measures $\muxi$ 
do not describe resonance eigenfunctions as good as previous approaches.
Altogether, our numerical results suggest that $\muxi$ is not the semiclassical 
limit measure, unless $\xi =0$ or $\xi = 1$.
Possible improvements could consider 
alternative combinations of $\munat$ and $\muinv$ or incorporate local 
Ehrenfest times instead of the fixed iteration number in Eq.~\eqref{EQ:muxi}.

Finally, it is straightforward to generalize the construction
of measures $\muxi$ to true time maps~\cite{AltPorTel2013b,AltPorTel2015}.
This would allow for a description of resonance eigenfunctions in billiards
with partial escape and thus in models of optical microcavities.
There the structure of resonance eigenfunctions has observable consequences
in the far-field emission of these cavities.

\begin{acknowledgments}
    We gratefully thank T. Becker, J. Keating, M. K\"orber,
    S. Nonnenmacher, M. Novaes,
    S. Prado, and M. Sieber for helpful comments and inspiring discussions.
    We thankfully acknowledge financial support
    through the Deutsche Forschungsgemeinschaft
    under Grant No. KE 537/5-1, from the IMPRS-MPSSE Dresden,
    from the Graduate Academy of TU Dresden, and from the University of Sydney 
    bridging Grant G199768.
\end{acknowledgments}

\section{Appendix}

\appendix

\section{Example: Standard map}\label{appendix.standardmap}

\subsection{Classical}\label{appendix.classical}

In order to investigate
the inverse measure $\muinv$
and the intermediate measures $\muxi$
numerically,
we consider the paradigmatic standard map \cite{Chi1979}.
It is given by the time-periodically driven Hamiltonian 
$H(q, p) = p^2/2 + \sum_n V(q) \delta(t - n)$, with
dimensionless coordinates $x \equiv (q, p)$ on phase-space
$\PS = [0, 1) \times [0,1)$ and kicking potential
$V(q) = \kappa/(4\pi^2) \cos(2\pi q)$.
We consider the half-kick mapping, which takes the form
$\Map(q, p) = (q + p^\ast, p^\ast - V'(q + p^\ast)/2)$ with
$p^\ast = p - V'(q)/2$.
All numerical results are computed for $\kappa = 10$,
for which the phase space is chaotic with no visible regular regions.
Escape is introduced by considering a phase-space region $\Omega$ (a leak) such that the reflectivity
$\refl(x) = \alphaOmega < 1$ for $x \in \Omega$
and $R(x) = 1$ for $x  \notin\Omega$.
This mimics the behavior of more realistic couplings
(e.g., Fresnel's law in optical microresonators). %
Numerical results are presented for a strip in $p$-direction on the
phase-space, $\Omega = (0.3, 0.6) \times [0, 1)$, with $\alphaOmega = 0.2$.

For the classical investigation we compute numerical approximations
of the measures $\muxi$.
First, we fix the number $n$ of time-steps for the approximation
$\tilmuxi^n$, see Eq.~\eqref{EQ:muxi}.
Secondly, we fix a set $X_c$ of $\sqrt{N_c} \times \sqrt{N_c}$
phase-space points on a grid.
The minimal distance $1/\sqrt{N_c}$ between two points defines the classical
resolution and should be much smaller than $\sqrt{h}$. Note
that for fixed $n$ and increasing $N_c$ we get better approximations
of $\tilmuxi^n$. 
For each grid point
$x\in X_c$ we compute the orbit $\{ M^k(x)\}_{k = -n}^{n-1}$.
which is used in Eq.~\eqref{EQ:muxi}.
There the integral $\tilmuxi^n(A)$ is given by the sum of all
contributions of points $x\in X_c\cap A$.
Finally we consider the approximate measure
$\muxi^n \equiv \frac{\tilmuxi^n}{\Vert \tilmuxi^n\Vert}$.
\begin{figure}[b!]
    \includegraphics[scale=1.]{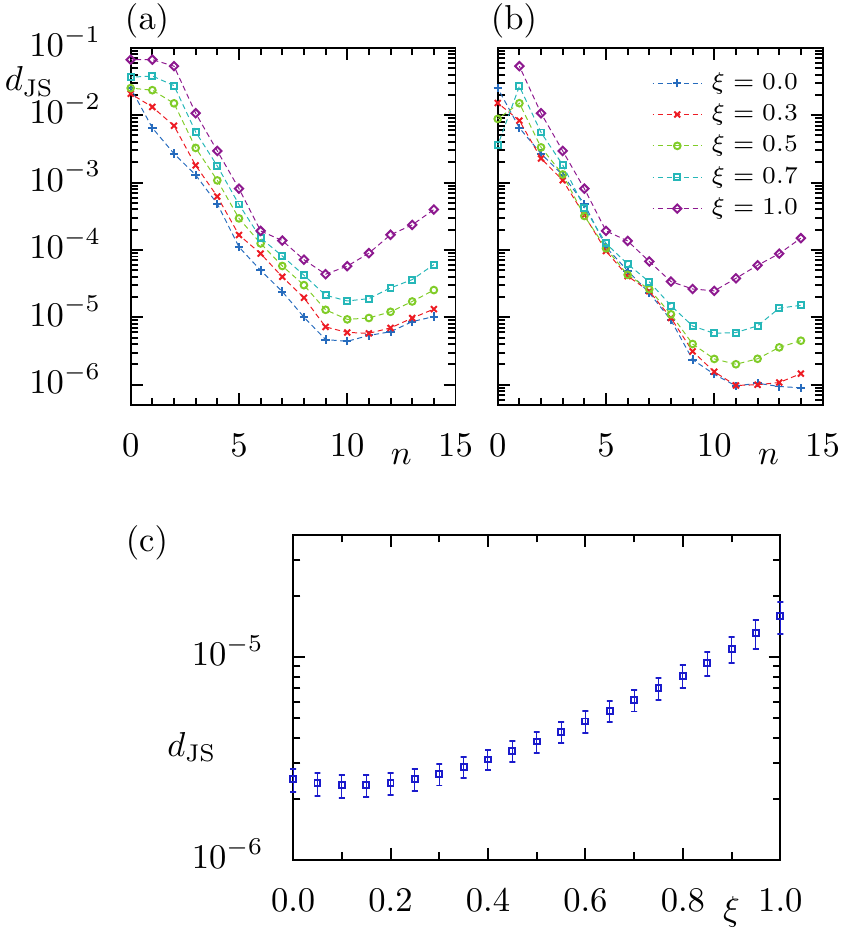}
    \caption{ %
        Jensen-Shannon divergence $\jsd(\epsilon=1/16)$ for
        estimating convergence, conditional invariance,
        and accuracy of classical construction for
        fixed number of initial conditions $N_c = 8192^2$.
        (a) Considered is $\jsd$ between $\muxi^n$ and
        its normalized iterate
        $\MMap\muxi^{n} / \Vert\MMap\muxi^n\Vert$
        over number of time-steps $n$ for $\xi \in\{0, 0.3, 0.5, 0.7, 1\}$.
        (b) Distance $\jsd(\muxi^n, \muxi^{n+1})$ between
        numerical constructions with increasing number of
        time-steps $n$ plotted over
        number of time-steps $n$ for same $\xi$ as in (a).
        (c) Average Jensen-Shannon divergence
        $\jsd(\muxi^n, \hat{\mu}_\xi^n)$ and
        standard deviation between
        $\muxi^n$ and $10$ different realizations of $\hat{\mu}^n_\xi$
        with random-uniform initial points on $\PS$ for fixed $n=8$.
    }
    \label{FIG:fig_app_cldist}
\end{figure}

In the following we report properties of the classical measures
regarding their conditional invariance, their convergence with $n$,
and the accuracy of the classical construction used in this paper.
This is illustrated in Fig.~\ref{FIG:fig_app_cldist}
for $N_c = 8192^2$ initial phase-space points.
Conditional invariance is investigated with the Jensen-Shannon
divergence $\jsd$ between approximation $\muxi^n$ and
its normalized iterate $\MMap\muxi^n / \Vert\MMap\muxi^n \Vert$,
see Fig.~\ref{FIG:fig_app_cldist}(a).
Increasing values of $n$ lead to a decreasing $\jsd$ for
all considered values of $\xi \in \{0, 0.3, 0.5, 0.7, 1\}$, up to
a maximal number $\tilde{n}$.
Thus $\muxi^n$ fulfills condition \eqref{EQ:cinv} with increasing $n$.
The limiting step $\tilde{n}$ is explained by the finite phase-space resolution
given by the fixed number of points $N_c$.

Secondly, we numerically show that $\jsd(\muxi^n, \muxi^{n+1})$ decreases
with $n$ for all considered $\xi$,
see Fig.~\ref{FIG:fig_app_cldist}(b).
Thus it indicates weak convergence of $\muxi^n$.
The same restrictions due to the finite number of points $N_c$ apply here.
Based on these results we use $\muxi^n$ with $n=8$ as an approximation
for $\muxi$ throughout the paper.
For these parameters we numerically obtain decay rates
$\gnat \approx 0.2165$ and $\ginv \approx 0.8820$.

Moreover, we calculate the Jensen-Shannon divergence
$\jsd$ between two different numerical approximations
$\muxi^n$ (defined above with $N_c$ grid-points) and $\hat{\mu}^n_\xi$
with $N_c$ random-uniform initial conditions.
This is needed to ensure that the distance between quantum and
classical measures is not influenced by classical fluctuations
due to the numerical construction.
We fix the number of time-steps $n=8$.
We estimate the classical error as the difference
between $\muxi^n$ and $\hat{\mu}^n_\xi$ for different realizations of
$\hat{\mu}^n_\xi$. 
The results are given in Fig.~\ref{FIG:fig_app_cldist}(c). Here we show
the average of the distances $\jsd(\muxi^n, \hat{\mu}^n_\xi)$ vs.\ $\xi$
and their standard-deviation for $10$ realizations.
This gives an estimate on the accuracy of the constructed and considered
measures $\muxi^n$. We observe that for $\xi = 0$ ($\gnat$) the accuracy
is nearly one magnitude smaller
than for $\xi = 1$ ($\ginv$), and that the dependence is continuous.
In all cases the errors are much smaller than the quantum-classical distances
investigated in Sec.~\ref{SEC:QuantitativeComparison}.

\subsection{Quantum}
\label{appendix.quantumopen}

The unitary quantum time-evolution in between two kicks is determined
by Floquet quantization \cite{BerBalTabVor1979,ChaShi1986}.
For the half-kick mapping it reads
\begin{align}
\QMapcls = \ue^{-\frac{\ui}{2\hbar}V(\hat{q}) }
\ue^{-\frac{\ui}{2\hbar}\hat{p}^2  }
\ue^{-\frac{\ui}{2\hbar}V(\hat{q}) },
\end{align}
where $h = 2\pi\hbar$ takes the role of an effective Planck's constant
due to dimensionless units $q$ and $p$. Considering periodic boundary 
conditions, i.e. dynamics on a torus, only discrete values $h = 1/N$
with $N\in\mathbb{N}$ are allowed.
The semiclassical limit is described by the limit $h\rightarrow 0$.
Due to the simple form of the reflectivity $\refl(x)$
the projective coupling operator \cite{NonSch2008}
takes the form
\begin{align}
\Qrefl = \proj_{\Omega^\uc} + \sqrt{\alphaOmega} \proj_\Omega.
\end{align}
Note, that the second term
$\sqrt{\alphaOmega}\proj_\Omega$ reduces the probability on 
$\Omega$ by a factor of $\alphaOmega$.
For $\alphaOmega < 1$ we obtain a subunitary time-evolution operator
$\QMap = \QMapcls\Qrefl$ for the system with escape.
Therefore all eigenvalues $\lambda = \ue^{\ui \theta - \gamma/2}$ have modulus
less than one, i.e.\ decay rates $\gamma > 0$.
We investigate the resonance eigenfunctions $\psi$ of $\QMap$
with decay rate $\gamma$ through
their Husimi phase-space distribution $\hus(x)$ \cite{Hus1940},
which is a smooth function. 
It is defined as the expectation
value of the state $\psi$ in a coherent state $|x\rangle$ centered at
$x\in\Gamma$,
$\hus_\psi(x) = h^{-1}|\langle x | \psi \rangle |^2$.
For a single eigenfunction $\psi$ this distribution shows strong quantum
fluctuations, illustrated in Fig.~\ref{FIG:fig1}(d-f).
A much clearer phase-space representation is obtained by
considering the average $\husavg$ over all $\hus_\psi$ where the
decay rate of $\psi$ is in the interval
$(\gamma - \Delta\gamma, \gamma + \Delta\gamma)$.

\section{Conditional invariance of $\muxi$}
\label{appendix:proof-ci}
Here we argue that $\muxi$ is a c-measure,
$\MMap\muxi = \ue^{-\gxi}\muxi$, Eq.~\eqref{EQ:cinv}, if
the considered closed map $\Map$ is uniformly hyperbolic.
Note that this condition is rather restrictive and is not satisfied
by the standard map.
The main idea is to use the local decomposition into stable and
unstable direction and thereby splitting the two-dimensional integrals
into two separate one-dimensional integrals.
For the natural measure we obtain that it is continuous in the
stable direction, while it is fractal along the unstable direction.
Locally we can write
$\munat(A) = \munat^u(A_1) \otimes \mu_\text{c}^s(A_2)$,
where $\munat^u$ and $\mu_c^s$ are fractal and continuous
measures on $\mathbb{R}$, respectively.
Note that this includes a local coordinate transformation from the
phase space to the tangent space,
including the Jacobian determinant in the integrals.
Time evolution $\MMap$ gives an additional factor $\refl$, see 
Eq.~\eqref{EQ:MMap0}, which
together with $\munat^u$ in the unstable direction leads
to the decay $\ue^{-\gnat}$.
On the other hand, for the inverse measure, we obtain fractality
along the stable direction only.
Similarly we can locally write
$\muinv(A) = \mu_\text{c}^u(A_1) \otimes \muinv^s(A_2)$ with
continuous $\mu_\text{c}^u$ and fractal $\muinv^s$ measures on $\mathbb{R}$.
Application of $\MMap$ gives the factor $\refl$, which
together with $\muinv^s$ in the stable direction
leads to the global decay $\ue^{-\ginv}$.

In this sense, the measures for intermediate decay rates
$\muxi$ can locally be understood as
$\muxi = \munat^u[\refl^{1-\xi}] \otimes \muinv^s[\refl^\xi]$,
where $\munat^u[\refl^{1-\xi}]$ is the natural measure obtained
for the map with escape $\refl^{1-\xi}$
and $\muinv^s[\refl^\xi]$ is the inverse measure obtained
for the map with escape $\refl^{\xi}$.
The additional factor $\refl$ from time-evolution,
Eq.~\eqref{EQ:MMap0},
can be split into $\refl^{1-\xi}$ and $\refl^\xi$.
The first leads to the decay rate $\gnat[\refl^{1-\xi}]$
corresponding to $\munat^u[\refl^{1-\xi}]$, while the second
gives $\ginv[\refl^\xi]$ corresponding to  $\muinv^s[\refl^\xi]$.
Thus the decay rate of the measures $\muxi$ is given
by Eq.~\eqref{EQ:GammaXi},
$\gxi = \gnat[\refl^{1-\xi}] + \ginv[\refl^\xi]$.

\section{Entropy of Husimi distributions}
\label{appendix.entropy}

We want to define the entropy of Husimi distributions of eigenfunctions
with decay rate $\gamma$.
Defining the probability measure $\mu_\hus(A) = \int_A \hus\, \ud\muL$
for each $\gamma$ we obtain the entropies $S(\epsilon; \mu_\hus)$,
Eq.~\eqref{EQ:Entropy}.
Because we are interested in the dependence on $\epsilon$, we
reduce fluctuations by considering the average entropy 
of eigenfunctions with decay rate in the interval
$(\gamma - \Delta \gamma, \gamma + \Delta \gamma)$.
This is given for fixed $\epsilon$ by
\begin{align}
 \langle S_\epsilon \rangle_\gamma &=
  \frac{1}{n_\gamma}
  \sum\limits_{|\gamma - \tilde{\gamma}| < \Delta\gamma}
  S(\epsilon; \mu_\hus) \\
  &= - \sum\limits_{A \in \mathcal{A}} \ 
 \frac{1}{n_\gamma}
 \sum\limits_{|\gamma - \tilde{\gamma}| < \Delta\gamma}
 p_{\tilde{\gamma},A} \log p_{\tilde{\gamma},A},%
\end{align}%
with number of eigenfunctions in the interval
$ n_\gamma = {\# \{ |\gamma - \tilde{\gamma}| < \Delta\gamma\}}$ 
and $p_{\tilde{\gamma},A} = \int_A \hus_\gamma\, \ud\muL$.
Fixing $\epsilon$ it is possible
to define an effective semiclassical limit as
$S_q(\epsilon;\gamma) \equiv
\lim_{h\rightarrow 0}\langle S_\epsilon \rangle_\gamma$ with fixed decay 
rate $\gamma$ and decreasing $h$.
The scaling of $S_q$ with $\epsilon$ eventually defines the
fractal dimension $\Dim $ of the quantum limit.
Numerically we have to deal
with finite values of $h$ and
effective entropies at this value.
We determine the fractal dimension
$\Dim$ of the quantum eigenfunctions
with decay rate $\gamma$
from the scaling of $\langle S\rangle_\gamma$ with $\epsilon$.
Note that in this case
Planck's scale $h$ defines a minimal resolution for $\epsilon$
below which there is no fractality and the trivial result $\Dim = 2$
is recovered.


\begin{thebibliography}{56}
\newcommand{\enquote}[1]{``#1''}
\providecommand{\url}[1]{\texttt{#1}}
\providecommand{\urlprefix}{URL }
\providecommand{\eprint}[2][]{\url{#2}}

\bibitem{BohGiaSch1984}
O.~Bohigas, M.~J. Giannoni, and C.~Schmit: \enquote{Characterization of chaotic
    quantum spectra and universality of level fluctuation laws},
\href{http://dx.doi.org/10.1103/PhysRevLett.52.1}{Phys.~Rev.~Lett.
    \textbf{52}, 1--4} (1984).

\bibitem{Ber1985}
M.~V. Berry: \enquote{Semiclassical theory of spectral rigidity},
\href{http://dx.doi.org/10.1098/rspa.1985.0078}{Proc.~R.~Soc.~Lon.~A
    \textbf{400}, 229--251} (1985).

\bibitem{SieRic2001}
M.~Sieber and K.~Richter: \enquote{Correlations between periodic orbits and
    their r\^ole in spectral statistics},
\href{http://dx.doi.org/10.1238/Physica.Topical.090a00128}{Phys.~Scripta
    \textbf{2001}, 128--133} (2001).

\bibitem{MueHeuBraHaaAlt2004}
S.~M\"uller, S.~Heusler, P.~Braun, F.~Haake, and A.~Altland:
\enquote{Semiclassical foundation of universality in quantum chaos},
\href{http://dx.doi.org/10.1103/PhysRevLett.93.014103}{Phys.~Rev.~Lett.
    \textbf{93}, 014103} (2004).

\bibitem{Vor1979}
A.~Voros: \enquote{Semi-classical ergodicity of quantum eigenstates in the
    {W}igner representation}, in G.~Casati and J.~Ford (editors)
\enquote{Stochastic Behavior in Classical and Quantum Hamiltonian Systems},
\href{http://dx.doi.org/10.1007/BFb0021732}{volume~93 of \emph{Lecture Notes
        in Physics}, 326--333, Springer Berlin Heidelberg, Berlin} (1979).

\bibitem{Ber1977b}
M.~V. Berry: \enquote{Regular and irregular semiclassical wavefunctions},
\href{http://dx.doi.org/10.1088/0305-4470/10/12/016}{J.~Phys.~A \textbf{10},
    2083--2091} (1977).

\bibitem{Ber1983}
M.~V. Berry: \enquote{Semiclassical mechanics of regular and irregular motion},
in G.~Iooss, R.~H.~G. Helleman, and R.~Stora (editors) \enquote{Comportement
    Chaotique des Syst{\`e}mes D{\'e}terministes --- Chaotic Behaviour of
    Deterministic Systems}, 171--271, {N}orth-{H}olland, {A}msterdam (1983).

\bibitem{Shn1974}
A.~I. Shnirelman: \enquote{Ergodic properties of eigenfunctions \rm (in
    {Russian})}, \href{http://mi.mathnet.ru/eng/umn4463}{Usp.~Math.~Nauk
    \textbf{29}, 181--182} (1974).

\bibitem{CdV1985}
Y.~{Colin de Verdi\`ere}: \enquote{Ergodicit\'e et fonctions propres du
    laplacien \rm (in {French})},
\href{http://projecteuclid.org/euclid.cmp/1104114465}{Commun.~Math.~Phys.
    \textbf{102}, 497--502} (1985).

\bibitem{Zel1987}
S.~Zelditch: \enquote{Uniform distribution of eigenfunctions on compact
    hyperbolic surfaces},
\href{http://dx.doi.org/10.1215/S0012-7094-87-05546-3}{Duke. Math. J.
    \textbf{55}, 919--941} (1987).

\bibitem{ZelZwo1996}
S.~Zelditch and M.~Zworski: \enquote{Ergodicity of eigenfunctions for ergodic
    billiards},
\href{http://projecteuclid.org/euclid.cmp/1104276097}{Commun.~Math.~Phys.
    \textbf{175}, 673--682} (1996).

\bibitem{NonVor1998}
S.~Nonnenmacher and A.~Voros: \enquote{Chaotic eigenfunctions in phase space},
\href{http://dx.doi.org/10.1023/A:1023080303171}{J.~Stat.~Phys. \textbf{92},
    431--518} (1998).

\bibitem{BaeSchSti1998}
A.~B\"acker, R.~Schubert, and P.~Stifter: \enquote{Rate of quantum ergodicity
    in {E}uclidean billiards},
\href{http://dx.doi.org/10.1103/PhysRevE.57.5425}{Phys.~Rev.~E \textbf{57},
    5425--5447} (1998), ; erratum ibid. {\bf 58} (1998) 5192.

\bibitem{Bie2001}
S.~De~Bi{\`e}vre: \enquote{Quantum chaos: a brief first visit}, in
S.~P{\'e}rez-Esteva and C.~Villegas-Blas (editors) \enquote{{Second {{Summer
                School}} in {{Analysis}} and {{Mathematical Physics}} 
                ({{Cuernavaca}},
        2000)}}, \href{http://dx.doi.org/10.1090/conm/289/04878}{Contemp. Math. 
        {\bf
        289}, 161--218, {Amer. Math. Soc., Providence, RI}} (2001).

\bibitem{CaoWie2015}
H.~Cao and J.~Wiersig: \enquote{Dielectric microcavities: Model systems for
    wave chaos and non-{Hermitian} physics},
\href{http://dx.doi.org/10.1103/RevModPhys.87.61}{Rev.~Mod.~Phys.
    \textbf{87}, 61--111} (2015).

\bibitem{MitRicWei2010}
G.~E. Mitchell, A.~Richter, and H.~A. Weidenm{\"u}ller: \enquote{Random
    matrices and chaos in nuclear physics: {N}uclear reactions},
\href{http://dx.doi.org/10.1103/RevModPhys.82.2845}{Rev.~Mod.~Phys.
    \textbf{82}, 2845--2901} (2010).

\bibitem{Sto1999}
H.-J. St\"ockmann\href{http://dx.doi.org/10.1017/CBO9780511524622}{:
    \emph{Quantum Chaos: {A}n {I}ntroduction}, Cambridge University Press,
    Cambridge} (1999).

\bibitem{LaiTel2011}
Y.-C. Lai and T.~T\'el: \emph{Transient Chaos: {C}omplex Dynamics on Finite
    Time Scales}, number 173 in Applied Mathematical Sciences, Springer Verlag,
New York, 1 edition (2011).

\bibitem{AltPorTel2013}
E.~G. Altmann, J.~S.~E. Portela, and T.~T\'el: \enquote{Leaking chaotic
    systems}, \href{http://dx.doi.org/10.1103/RevModPhys.85.869}{Rev.~Mod.~Phys.
    \textbf{85}, 869--918} (2013).

\bibitem{Gas2014b}
P.~Gaspard: \enquote{Quantum chaotic scattering},
\href{http://dx.doi.org/10.4249/scholarpedia.9806}{{S}cholarpedia
    \textbf{9(6)}, 9806} (2014).

\bibitem{CasMasShe1999b}
G.~Casati, G.~Maspero, and D.~L. Shepelyansky: \enquote{Quantum fractal
    eigenstates},
\href{http://dx.doi.org/10.1016/S0167-2789(98)00265-6}{Physica~D
    \textbf{131}, 311--316} (1999).

\bibitem{KeaNovPraSie2006}
J.~P. Keating, M.~Novaes, S.~D. Prado, and M.~Sieber: \enquote{Semiclassical
    structure of chaotic resonance eigenfunctions},
\href{http://dx.doi.org/10.1103/PhysRevLett.97.150406}{Phys.~Rev.~Lett.
    \textbf{97}, 150406} (2006).

\bibitem{NonRub2007}
S.~Nonnenmacher and M.~Rubin: \enquote{Resonant eigenstates for a quantized
    chaotic system},
\href{http://dx.doi.org/10.1088/0951-7715/20/6/004}{Nonlinearity \textbf{20},
    1387--1420} (2007).

\bibitem{BorGuaShe1991}
F.~Borgonovi, I.~Guarneri, and D.~L. Shepelyansky: \enquote{Statistics of
    quantum lifetimes in a classically chaotic system},
\href{http://dx.doi.org/10.1103/PhysRevA.43.4517}{Phys.~Rev.~A \textbf{43},
    4517--4520} (1991).

\bibitem{SchTwo2004}
H.~Schomerus and J.~Tworzyd{\l}o: \enquote{Quantum-to-classical crossover of
    quasibound states in open quantum systems},
\href{http://dx.doi.org/10.1103/PhysRevLett.93.154102}{Phys.~Rev.~Lett.
    \textbf{93}, 154102} (2004).

\bibitem{KeaNonNovSie2008}
J.~P. Keating, S.~Nonnenmacher, M.~Novaes, and M.~Sieber: \enquote{On the
    resonance eigenstates of an open quantum baker map},
\href{http://stacks.iop.org/0951-7715/21/i=11/a=007}{Nonlinearity
    \textbf{21}, 2591--2624} (2008).

\bibitem{ErmCarSar2009}
L.~Ermann, G.~G. Carlo, and M.~Saraceno: \enquote{Localization of resonance
    eigenfunctions on quantum repellers},
\href{http://dx.doi.org/10.1103/PhysRevLett.103.054102}{Phys.~Rev.~Lett.
    \textbf{103}, 054102} (2009).

\bibitem{ClaKoeBaeKet2018}
K.~Clau\ss, M.~J. K\"orber, A.~B\"acker, and R.~Ketzmerick: \enquote{Resonance
    eigenfunction hypothesis for chaotic systems},
\href{http://dx.doi.org/10.1103/PhysRevLett.121.074101}{Phys.~Rev.~Lett.
    \textbf{121}, 074101} (2018).

\bibitem{BilGarGeoGir2019}
A.~M. Bilen, I.~{Garc{\'i}a-Mata}, B.~Georgeot, and O.~Giraud:
\enquote{Multifractality of open quantum systems},
\href{http://dx.doi.org/10.1103/PhysRevE.100.032223}{Phys.~Rev.~E
    \textbf{100}, 032223} (2019).

\bibitem{GasRic1989a}
P.~Gaspard and S.~A. Rice: \enquote{Scattering from a classically chaotic
    repellor}, \href{http://dx.doi.org/10.1063/1.456017}{J.~Chem.~Phys.
    \textbf{90}, 2225--2241} (1989).

\bibitem{PiaYor1979}
G.~Pianigiani and J.~A. Yorke: \enquote{Expanding maps on sets which are almost
    invariant: {D}ecay and chaos},
\href{http://dx.doi.org/10.2307/1998093}{Trans.~Amer.~Math.~Soc.
    \textbf{252}, 351--366} (1979).

\bibitem{DemYou2006}
M.~F. Demers and L.-S. Young: \enquote{Escape rates and conditionally invariant
    measures}, \href{http://dx.doi.org/10.1088/0951-7715/19/2/008}{Nonlinearity
    \textbf{19}, 377--397} (2006).

\bibitem{Sjo1990}
J.~Sj{\"o}strand: \enquote{Geometric bounds on the density of resonances for
    semiclassical problems},
\href{http://dx.doi.org/10.1215/S0012-7094-90-06001-6}{Duke Math.~J.
    \textbf{60}, 1--57} (1990).

\bibitem{Lin2002}
K.~K. Lin: \enquote{Numerical study of quantum resonances in chaotic
    scattering}, 
    \href{http://dx.doi.org/10.1006/jcph.2001.6986}{J.~Comput.~Phys.
    \textbf{176}, 295--329} (2002).

\bibitem{LuSriZwo2003}
W.~T. Lu, S.~Sridhar, and M.~Zworski: \enquote{Fractal {W}eyl laws for chaotic
    open systems},
\href{http://dx.doi.org/10.1103/PhysRevLett.91.154101}{Phys.~Rev.~Lett.
    \textbf{91}, 154101} (2003).

\bibitem{RamPraBorFar2009}
J.~A. Ramilowski, S.~D. Prado, F.~Borondo, and D.~Farrelly: \enquote{Fractal
    {W}eyl law behavior in an open {H}amiltonian system},
\href{http://dx.doi.org/10.1103/PhysRevE.80.055201}{Phys.~Rev.~E \textbf{80},
    055201(R)} (2009).

\bibitem{EbeMaiWun2010}
A.~Ebersp\"acher, J.~Main, and G.~Wunner: \enquote{Fractal {W}eyl law for
    three-dimensional chaotic hard-sphere scattering systems},
\href{http://dx.doi.org/10.1103/PhysRevE.82.046201}{Phys.~Rev.~E \textbf{82},
    046201} (2010).

\bibitem{ErmShe2010}
L.~Ermann and D.~L. Shepelyansky: \enquote{Ulam method and fractal {W}eyl law
    for {P}erron-{F}robenius operators},
\href{http://dx.doi.org/10.1140/epjb/e2010-00144-0}{Eur.~Phys.~J.~B
    \textbf{75}, 299--304} (2010).

\bibitem{NonSjoZwo2014}
S.~Nonnenmacher, J.~Sj\"ostrand, and M.~Zworski: \enquote{Fractal {W}eyl law
    for open quantum chaotic maps},
\href{http://dx.doi.org/10.4007/annals.2014.179.1.3}{Annals of Mathematics
    \textbf{179}, 179--251} (2014).

\bibitem{LeeRimRyuKwoChoKim2004}
S.-Y. Lee, S.~Rim, J.-W. Ryu, T.-Y. Kwon, M.~Choi, and C.-M. Kim:
\enquote{Quasiscarred resonances in a spiral-shaped microcavity},
\href{http://dx.doi.org/10.1103/PhysRevLett.93.164102}{Phys.~Rev.~Lett.
    \textbf{93}, 164102} (2004).

\bibitem{WieHen2008}
J.~Wiersig and M.~Hentschel: \enquote{Combining directional light output and
    ultralow loss in deformed microdisks},
\href{http://dx.doi.org/10.1103/PhysRevLett.100.033901}{Phys.~Rev.~Lett.
    \textbf{100}, 033901} (2008).

\bibitem{WieMai2008}
J.~Wiersig and J.~Main: \enquote{Fractal {W}eyl law for chaotic microcavities:
    Fresnel's laws imply multifractal scattering},
\href{http://dx.doi.org/10.1103/PhysRevE.77.036205}{Phys.~Rev.~E \textbf{77},
    036205} (2008).

\bibitem{ShiWieCao2011}
J.-B. Shim, J.~Wiersig, and H.~Cao: \enquote{Whispering gallery modes formed by
    partial barriers in ultrasmall deformed microdisks},
\href{http://dx.doi.org/10.1103/PhysRevE.84.035202}{Phys.~Rev.~E \textbf{84},
    035202(R)} (2011).

\bibitem{ShiHarFukHenSasNar2010}
S.~Shinohara, T.~Harayama, T.~Fukushima, M.~Hentschel, T.~Sasaki, and E.~E.
Narimanov: \enquote{Chaos-assisted directional light emission from
    microcavity lasers},
\href{http://dx.doi.org/10.1103/PhysRevLett.104.163902}{Phys.~Rev.~Lett.
    \textbf{104}, 163902} (2010).

\bibitem{AltPorTel2013b}
E.~G. Altmann, J.~S.~E. Portela, and T.~T\'el: \enquote{Chaotic systems with
    absorption},
\href{http://dx.doi.org/10.1103/PhysRevLett.111.144101}{Phys.~Rev.~Lett.
    \textbf{111}, 144101} (2013).

\bibitem{HarShi2015}
T.~Harayama and S.~Shinohara: \enquote{Ray-wave correspondence in chaotic
    dielectric billiards},
\href{http://dx.doi.org/10.1103/PhysRevE.92.042916}{Phys.~Rev.~E \textbf{92},
    042916} (2015).

\bibitem{AltPorTel2015}
E.~G. Altmann, J.~S.~E. Portela, and T.~T\'el: \enquote{Chaotic explosions},
\href{http://dx.doi.org/10.1209/0295-5075/109/30003}{EPL \textbf{109}, 30003}
(2015).

\bibitem{KulWie2016}
J.~Kullig and J.~Wiersig: \enquote{Frobenius--{Perron} eigenstates in deformed
    microdisk cavities: non-{Hermitian} physics and asymmetric backscattering in
    ray dynamics}, \href{http://dx.doi.org/10.1088/1367-2630/18/1/015005}{New J.
    Phys. \textbf{18}, 015005} (2016).

\bibitem{NonSch2008}
S.~Nonnenmacher and E.~Schenck: \enquote{Resonance distribution in open quantum
    chaotic systems},
\href{http://dx.doi.org/10.1103/PhysRevE.78.045202}{Phys.~Rev.~E \textbf{78},
    045202(R)} (2008).

\bibitem{PotWeiBarKuhStoZwo2012}
A.~Potzuweit, T.~Weich, S.~Barkhofen, U.~Kuhl, H.-J. St\"ockmann, and
M.~Zworski: \enquote{Weyl asymptotics: {F}rom closed to open systems},
\href{http://dx.doi.org/10.1103/PhysRevE.86.066205}{Phys.~Rev.~E \textbf{86},
    066205} (2012).

\bibitem{GutOsi2015}
B.~Gutkin and V.~A. Osipov: \enquote{Universality in spectral statistics of
    open quantum graphs},
\href{http://dx.doi.org/10.1103/PhysRevE.91.060901}{Phys.~Rev.~E \textbf{91},
    060901(R)} (2015).

\bibitem{SchAlt2015}
M.~Sch\"onwetter and E.~G. Altmann: \enquote{Quantum signatures of classical
    multifractal measures},
\href{http://dx.doi.org/10.1103/PhysRevE.91.012919}{Phys.~Rev.~E \textbf{91},
    012919} (2015).

\bibitem{KoeMicBaeKet2013}
M.~J. K\"orber, M.~Michler, A.~B\"acker, and R.~Ketzmerick:
\enquote{Hierarchical fractal {W}eyl laws for chaotic resonance states in
    open mixed systems},
\href{http://dx.doi.org/10.1103/PhysRevLett.111.114102}{Phys.~Rev.~Lett.
    \textbf{111}, 114102} (2013).

\bibitem{CarBenBor2016}
G.~G. Carlo, R.~M. Benito, and F.~Borondo: \enquote{Theory of short periodic
    orbits for partially open quantum maps},
\href{http://dx.doi.org/10.1103/PhysRevE.94.012222}{Phys.~Rev.~E \textbf{94},
    012222} (2016).

\bibitem{PraCarBenBor2018}
C.~A. Prado, G.~G. Carlo, R.~M. Benito, and F.~Borondo: \enquote{Role of short
    periodic orbits in quantum maps with continuous openings},
\href{http://dx.doi.org/10.1103/PhysRevE.97.042211}{Phys.~Rev.~E \textbf{97},
    042211} (2018).

\bibitem{LipRyuLeeKim2012}
D.~Lippolis, J.-W. Ryu, S.-Y. Lee, and S.~W. Kim: \enquote{On-manifold
    localization in open quantum maps},
\href{http://dx.doi.org/10.1103/PhysRevE.86.066213}{Phys.~Rev.~E \textbf{86},
    066213} (2012).

\bibitem{LipRyuKim2015}
D.~Lippolis, J.-W. Ryu, and S.~W. Kim: \enquote{Localization in chaotic systems
    with a single-channel opening},
\href{http://dx.doi.org/10.1103/PhysRevE.92.012921}{Phys.~Rev.~E \textbf{92},
    012921} (2015).

\bibitem{FyoSom2000}
Y.~V. Fyodorov and H.-J. Sommers: \enquote{Spectra of random contractions and
    scattering theory for discrete-time systems},
\href{http://dx.doi.org/10.1134/1.1335121}{J.~Exp.~Theor.~Phys.~Lett.
    \textbf{72}, 422--426} (2000).

\bibitem{KeaNovSch2008}
J.~P. Keating, M.~Novaes, and H.~Schomerus: \enquote{Model for chaotic
    dielectric microresonators},
\href{http://dx.doi.org/10.1103/PhysRevA.77.013834}{Phys.~Rev.~A \textbf{77},
    013834} (2008).

\bibitem{She2008}
D.~L. Shepelyansky: \enquote{Fractal {W}eyl law for quantum fractal
    eigenstates},
\href{http://dx.doi.org/10.1103/PhysRevE.77.015202}{Phys.~Rev.~E \textbf{77},
    015202(R)} (2008).

\bibitem{Hus1940}
K.~Husimi: \enquote{Some formal properties of the density matrix}, Proc. Phys.
Math. Soc. Jpn. \textbf{22}, 264--314 (1940).

\bibitem{CheMar1997a}
N.~Chernov and R.~Markarian: \enquote{Ergodic properties of {A}nosov maps with
    rectangular holes}, \href{http://dx.doi.org/10.1007/BF01233395}{Bol. Soc.
    Bras. Mat \textbf{28}, 271--314} (1997).

\bibitem{CheMarTro2000}
N.~Chernov, R.~Markarian, and S.~Troubetzkoy: \enquote{Invariant measures for
    {A}nosov maps with small holes},
\href{http://dx.doi.org/10.1017/S0143385700000560}{Ergodic Theory
    Dynam.~Systems \textbf{20}, 1007--1044} (2000).

\bibitem{Ula1960}
S.~M. Ulam: \enquote{A collection of mathematical problems}, in L.~Bers,
R.~Courant, and J.~J. Stoker (editors) \enquote{Interscience tracts in pure
    and applied Mathematics}, volume~8, 73--75, Interscience, New York (1960).

\bibitem{GroBerCarRomOliSta2002}
I.~Grosse, P.~{Bernaola-Galv\'an}, P.~Carpena, R.~{Rom\'an-Rold\'an},
J.~Oliver, and H.~E. Stanley: \enquote{Analysis of symbolic sequences using
    the {J}ensen-{S}hannon divergence},
\href{http://dx.doi.org/10.1103/PhysRevE.65.041905}{Phys.~Rev.~E \textbf{65},
    041905} (2002).

\bibitem{KoeBaeKet2015}
M.~J. K\"orber, A.~B\"acker, and R.~Ketzmerick: \enquote{Localization of
    chaotic resonance states due to a partial transport barrier},
\href{http://dx.doi.org/10.1103/PhysRevLett.115.254101}{Phys.~Rev.~Lett.
    \textbf{115}, 254101} (2015).

\bibitem{Chi1979}
B.~V. {Chirikov}: \enquote{{A universal instability of many-dimensional
        oscillator systems}},
\href{http://dx.doi.org/10.1016/0370-1573(79)90023-1}{Phys.~Rep. \textbf{52},
    263--379} (1979).

\bibitem{BerBalTabVor1979}
M.~V. Berry, N.~L. Balazs, M.~Tabor, and A.~Voros: \enquote{Quantum maps},
\href{http://dx.doi.org/10.1016/0003-4916(79)90296-3}{Ann.~Phys.~(N.Y.)
    \textbf{122}, 26--63} (1979).

\bibitem{ChaShi1986}
S.-J. Chang and K.-J. Shi: \enquote{Evolution and exact eigenstates of a
    resonant quantum system},
\href{http://dx.doi.org/10.1103/PhysRevA.34.7}{Phys.~Rev.~A \textbf{34},
    7--22} (1986).

\end{thebibliography}
\end{document}